\documentclass[submission,copyright,creativecommons]{eptcs}
\providecommand{\event}{LSFA 2026} % Name of the event you are submitting to

\usepackage{iftex}

\ifpdf
  \usepackage{underscore}         % Only needed if you use pdflatex.
  \usepackage[T1]{fontenc}        % Recommended with pdflatex
\else
  \usepackage{breakurl}           % Not needed if you use pdflatex only.
\fi

\title{Templates in Rewriting Induction}
\author{
    Kasper Hagens 
    \institute{Radboud University, Nijmegen}
    \email{kasper.hagens@ru.nl}
    \and
    Cynthia Kop \thanks{Both authors are supported by the NWO VIDI project VI.Vidi.193.075.}
    \institute{Radboud University, Nijmegen}
    \email{c.kop@cs.ru.nl}
}
\def\titlerunning{Templates in Rewriting Induction}
\def\authorrunning{K. Hagens \& C. Kop}
\usepackage{enumitem}
\usepackage{amsmath,amssymb, mathtools}
\usepackage{amsthm}
\theoremstyle{plain}      
\newtheorem{theorem}{Theorem}[section]  
\newtheorem{lemma}[theorem]{Lemma} 
\theoremstyle{definition}  
\newtheorem{definition}[theorem]{Definition}
\newtheorem{example}[theorem]{Example}
\newtheorem{remark}[theorem]{Remark}

\usepackage{cleveref} 
\Crefname{theorem}{Theorem}{Theorems}
\Crefname{lemma}{Lemma}{Lemmas}
\Crefname{definition}{Definition}{Definitions}
\Crefname{example}{Example}{Examples}
\Crefname{remark}{Remark}{Remarks}

\crefname{theorem}{theorem}{theorems}
\crefname{lemma}{lemma}{lemmas}
\crefname{definition}{definition}{definitions}
\crefname{example}{example}{examples}
\crefname{remark}{remark}{remarks}

\usepackage{upgreek} 
\usepackage{bm}
\usepackage{scalerel} 
\usepackage{bussproofs}
\usepackage{subcaption}

\usepackage{xcolor-material}

\usepackage{listings}
\lstdefinestyle{customc}{
  belowcaptionskip=1\baselineskip,
  breaklines=true,
  frame=L,
  xleftmargin=\parindent,
  language=C,
  showstringspaces=false,
  basicstyle=\footnotesize\ttfamily,
  keywordstyle=\bfseries\color{blue!90!black},
  commentstyle=\itshape\color{purple},
  identifierstyle=\color{black},
  stringstyle=\color{orange},
}

\newcommand{\myparagraph}[1]{\medskip\noindent\textbf{#1.}}

\newcommand{\cora}{\textsf{Cora}}
\newcommand{\lcstrs}{LCSTRS}
\newcommand{\lcstrss}{LCSTRSs}

\newcommand{\Sorts}{\mathbb{S}}
\newcommand{\thSorts}{\Sorts_{theory}}
\newcommand{\Types}{\mathcal{T}}
\newcommand{\thTypes}{\Types_{theory}}

\newcommand{\typeInterpret}[1]{\mathcal{I}_{#1}}

\newcommand{\arrtype}{\to}

\newcommand{\atype}{\sigma}
\newcommand{\btype}{\tau}
\newcommand{\asort}{\iota}
\newcommand{\bsort}{\kappa}
\newcommand{\arrz}{\to}

\renewcommand{\int}{\mathsf{int}}
\newcommand{\bool}{\mathsf{bool}}

\newcommand{\strue}{\mathtt{true}}
\newcommand{\sfalse}{\mathtt{false}}
\newcommand{\true}{\top}
\newcommand{\false}{\bot}

\newcommand{\Sig}{\Sigma}
\newcommand{\thSig}{\Sig_{theory}}
\newcommand{\termsSig}{\Sig_{terms}}
\newcommand{\Var}{\mathcal{V}}
\newcommand{\Vars}[1]{Var(#1)}
\newcommand{\Val}{\mathcal{V}\!\textit{al}}
\newcommand{\Terms}{T(\Sig, \Var)}

\newcommand{\typeof}{\mathit{typeof}}
\newcommand{\supterm}{\triangleright}
\newcommand{\domain}{\mathit{dom}}
\newcommand{\arity}{\mathit{ar}}
\newcommand{\termInterpret}[1]{[\![ #1 ]\!]}
\newcommand{\prefix}[1]{\mathsf{[#1]}}
\newcommand{\symb}[1]{\mathsf{#1}}

\newcommand{\sterm}{\varsigma}
\newcommand{\afun}{\mathsf{f}}
\newcommand{\tterm}{\uptau}
\newcommand{\avar}{x}
\newcommand{\bvar}{y}
\newcommand{\cvar}{z}

\newcommand{\rules}{\mathcal{R}}
\newcommand{\calcrules}{\rules_{calc}}
\newcommand{\myrule}{\ell \to r \ [\varphi]}

\newcommand{\rw}{\to_{\rules}}
\newcommand{\rweq}[3]{{#1} \mathrel{\leftrightarrow^{*}_{#3}} {#2}}
\newcommand{\openconv}[4]{{#1} \mathrel{\xleftrightarrow{{\scriptscriptstyle #3\succeq}}^*_{#4}}{#2}}
\newcommand{\rconv}[3]{\openconv{#1}{#2}{#3}{\rules}}

\newcommand{\eq}[2]{\textbf{(#1}_#2\textbf{)}}
\newcommand{\eqs}{\mathcal{E}}
\newcommand{\hs}{\mathcal{H}}
\newcommand{\vdashc}{%
  \mathbin{%
    \rlap{%
      \raisebox{1.1ex}{%
        \hspace{0.5em}%
      }%
    }%
    \vdash%
  }%
}

\newcommand{\factRD}{\mathsf{facRD}}
\newcommand{\factRU}{\mathsf{facRU}}
\newcommand{\factTU}{\mathsf{facTU}}
\newcommand{\factTD}{\mathsf{facTD}}

\newcommand{\funfactTU}{\mathsf{funfacTU}}
\newcommand{\funfactTD}{\mathsf{funfacTD}}
\newcommand{\downloop}{\mathsf{d}}
\newcommand{\uploop}{\mathsf{u}}
\newcommand{\uprec}{\mathsf{U}}
\newcommand{\recR}{\mathsf{R}}

\newcommand{\iter}{\mathsf{T}}

\newcommand{\tutemp}{T^{+}_{\funccont, \upterm}}
\newcommand{\tdtemp}{T^{-}_{\funccont, \lowterm}}
\newcommand{\rutemp}{R^{+}_{\funccont, \upterm, \baseterm}}
\newcommand{\rdtemp}{R^{-}_{\funccont, \lowterm, \baseterm}}
\newcommand{\upterm}{u}
\newcommand{\lowterm}{l}
\newcommand{\funcvar}{f}

\newcommand{\funccont}{F}
\newcommand{\baseterm}{b}
\newcommand{\ivar}{i}
\newcommand{\accvar}{a}
\newcommand{\upvar}{y}
\newcommand{\lowvar}{x}
\newcommand{\basevar}{z}
\newcommand{\ff}[2]{\funccont(#1,#2)}
\newcommand{\Qsymb}{\underline{\symb{F}}}

\newcommand{\rulesrec}{\rules_{\mathtt{rec}}}

\newcommand{\tub}{(\tau^\uparrow_b)}
\newcommand{\tur}{(\tau^\uparrow_r)}
\newcommand{\tdb}{(\tau^\downarrow_b)}
\newcommand{\tdr}{(\tau^\downarrow_r)}
\newcommand{\rub}{(\rho^\uparrow_b)}
\newcommand{\rur}{(\rho^\uparrow_r)}
\newcommand{\rdb}{(\rho^\downarrow_b)}
\newcommand{\rdr}{(\rho^\downarrow_r)}
\newcommand{\tailup}{\symb{tailup}}
\newcommand{\taildown}{\symb{taildn}}
\newcommand{\recup}{\symb{recup}}
\newcommand{\recdown}{\symb{recdn}}

\newcommand{\TURDlem}{\textbf{(L}_1\textbf{)}}
\newcommand{\TDRUlem}{\textbf{(L}_2\textbf{)}}

\newcommand{\lemref}[1]{Lemma~\ref{lem:#1}}

\newcommand{\exaref}[1]{Example~\ref{ex:#1}}

\begin{document}
\maketitle

\begin{abstract}
Rewriting Induction (RI) is a formal system in term rewriting to establish program equivalence.   
The recently defined Bounded RI for higher-order Logically Constrained Term Rewriting Systems
(\lcstrss) yields a convenient proof system for analyzing real programming code.

A practical challenge in RI is the automatic generation of induction hypotheses, called lemmas. 
While various lemma generation techniques exist for plain term rewriting, there are much fewer
that consider the intricacies brought on by calculations or constraints.
Taking advantage of recent developments in \emph{higher-order} RI, we here present a new approach
based on \emph{templates}, which operates by recognising typical programming constructs as
instances of higher-order functions.
While templates have been used as a stand-alone method to justify the correctness of program transformations, we here consider their integration in Bounded RI to obtain a complementary lemma generation heuristic.
This allows us to prove equivalences that were previously out of reach. 
\end{abstract}

\section{Introduction}
Rewriting Induction (RI) is a proof system for establishing inductive theorems via term rewriting.
Inductive theorems are used in formal verification to model program equivalence, based on the interpretation that two programs are equivalent if they exhibit the same input-output behavior.
For a rewrite system $\rules$, inductive theorems are defined by means of the \emph{convertibility} relation $\rweq{}{}{\rules}$, referring to the smallest equivalence relation that contains $\rules$ and which is closed under application of substitutions and contexts. 
An equation $s \approx t$ is an inductive theorem if it is ground convertible, meaning that $\rweq{s \gamma}{t \gamma}{\rules}$ for every \emph{ground} substitution $\gamma$ (a substitution that only substitutes \emph{variable-free} terms). 
For example, let 
\begin{center}
\(
\rules
=\{
\begin{aligned}
\symb{add}(\symb{0}, y)
\to 
y,\  
\symb{add}(\symb{s}(x), y)
\to 
\symb{s}(\symb{add}(x, y)) 
\end{aligned}
\}
\)
\end{center}
We claim that $\symb{add}(x, \symb{s}(\symb{0})) \approx \symb{s}(x)$ is an inductive theorem, which means that $\rweq{\symb{add}(x, \symb{s}(\symb{0})) \gamma}{\symb{s}(x) \gamma}{\rules}$ for all ground substitutions $\gamma$.
For concrete ground substitutions this is an easy property to check. 
For example, if $\gamma = [x :=\symb{s}(\symb{0})]$ then convertibility is witnessed by 
the reduction
\begin{center}
\(
\symb{add}(\symb{s}(\symb{0}), \symb{s}(\symb{0})) \to \symb{s}(\symb{add}(\symb{0}, \symb{s}(\symb{0}))) \to \symb{s}(\symb{s}(\symb{0}))
\)
\end{center}
To prove convertibility for an arbitrary ground substitution $\gamma$, we may use well-founded induction.
If $\rules$ is terminating then $\rw$ is a well-founded ordering, enabling the use of well-founded induction in RI.

\myparagraph{Constrained rewriting}
The RI system has been adapted to constrained rewriting~\cite{fal:kap:12,fuh:kop:nis:17,nak:nis:kus:sak:sak:10}, and recently to higher-order constrained
rewriting~\cite{hag:kop:24,hag:kop:26} using Logically Constrained Simply typed Term Rewriting Systems (\lcstrss). 
This formalism closely relate to real 
programming and therefore have a natural place in the larger toolbox for program verification. 
Programs are represented by term rewriting systems, and equivalence is modeled by inductive theorems.
Rewrite rules have a shape $s \to t\ [\varphi]$ where the boolean constraint $\varphi$ acts as a guard, in order to manage control flow over primitive data structures, such as integers.
Constraints are \emph{theory terms}: built from variables and \emph{theory symbols} (e.g. $+$ and $*$). 
Ground instances of theory terms have a semantical interpretation. 
For example, addition on integers is interpreted as addition on $\mathbb{Z}$. 
So for a theory term $t:=(x+y \ge \symb{5})$ and ground substitution $\gamma = [x:=\symb{2}, y:=\symb{3}]$
the instance $t\gamma$ has the semantical interpretation $\termInterpret{t \gamma} =
\termInterpret{\symb{2}+\symb{3} \ge \symb{5}}=(2+3 \ge 5) = \true$. 

\begin{figure}[t]
\begin{subfigure}{0.45\textwidth}
\begin{lstlisting}[language=C, style=customc]
int factTU(int x){
int a = 1; int i = 1;
while (i<=x){
    a = i*a; i = i+1;}
return a; }
\end{lstlisting}
\caption{iterative upward \phantom{WWWWWWW}}
\end{subfigure}
\begin{subfigure}{0.45\textwidth}
\begin{lstlisting}[language=C, style=customc]
int factTD(int x){
int a = 1;
while (x>0){
    a = a*x; x = x-1;}
return a; }
\end{lstlisting}
\caption{iterative downward \phantom{WWWWWW}}
\end{subfigure}

\medskip
\begin{subfigure}{0.45\textwidth}
\begin{lstlisting}[language=C, style=customc]
int factRD(int x){
if (x > 0) 
    return(x*factRD(x-1));
else 
    return 1; 
} 
\end{lstlisting}
\caption{recursive downward \phantom{WWWWWWW}}
\end{subfigure}
\begin{subfigure}{0.45\textwidth}
\begin{lstlisting}[language=C, style=customc]
int factRU(int x) = R(1, x);
int R(int i, int x){
if (i<x) 
    return(i*R(i+1, x));
else 
    return x; }
\end{lstlisting}
\caption{recursive upward \phantom{WWWWWWWWW}}
\end{subfigure}
\caption{Four implementations of the factorial function}
\label{fig:factcode}
\end{figure}

\begin{figure}[t]
\begin{subfigure}{0.45\textwidth}
\(\boxed{
\begin{aligned} 
&
\factTU\ x
\to 
\uploop\ x\ \symb{1}\ \symb{1}
\\
&
\uploop\ x\ i\ a 
\to 
a 
&
[i > x]
\\ 
&
\uploop\ x\ i\ a
\to 
\uploop\ x\ (i+\symb{1})\ (i*a)
&
[i \le x]
\end{aligned}}
\)
\caption{Tail recursive Upward (TU) \phantom{WWWWWWW}}
\label{fig:TU}
\end{subfigure}
\begin{subfigure}{0.45\textwidth}
\(\boxed{
\begin{aligned} 
&
\factTD\ x
\to 
\downloop\ x\ \symb{1}
\\
&
\downloop\ x\ a
\to 
a
&
[x \le \symb{0}]
\\
&
\downloop\ x\ a
\to 
\downloop\ (x-\symb{1})\ (a*x)
&
[x>\symb{0}]
\end{aligned}}
\)

\caption{Tail recursive Downward (TD) \phantom{WWWWWW}}
\label{fig:TD}
\end{subfigure}
\medskip

\begin{subfigure}{0.45\textwidth}
\(\boxed{
\begin{aligned}
\\
\factRD\ x
&
\to 
\symb{1}
&
[x \le \symb{1}]
\\
\factRD\ x
&
\to 
x*(\factRD\ (x-\symb{1}))
&
[x>\symb{1}]
\end{aligned}}
\)
\caption{Recursive Downward (RD) \phantom{WWWWWWW}}
\label{fig:RD}
\end{subfigure}
\begin{subfigure}{0.45\textwidth}
\(\boxed{
\begin{aligned} 
&
\factRU\ x
\to 
\uprec\ \symb{1}\ x
\\
&
\uprec\ i\ x
\to 
x
&
[i > x-\symb{1}]
\\
&
\uprec\ i\ x
\to 
i*(\uprec\ (i+\symb{1})\ x)
&
[i \le x-\symb{1}]
\end{aligned}}
\)
\caption{Recursive Upward (RU) \phantom{WWWWWWWWW}}
\label{fig:RU}
\end{subfigure}
\caption{\lcstrs\ translations of the implementations in \cref{fig:factcode}}
\label{fig:factlcstrs}
\end{figure}
We consider four implementations of the factorial function
in \cref{fig:factcode}, and represent them as Logically Constrained
Simply-typed Term Rewriting Systems in \cref{fig:factlcstrs}.
These implementations do not necessarily agree on inputs $x <1$, but if we restrict to $x \ge 1$ they all compute $x \mapsto \prod_{i=1}^{x} i$. 
Such a restriction is easily expressed in constrained rewriting, where also equations may be equipped with constraints, so as
to limit interest to
substitutions that satisfy the constraint.
For example, $\factTU\ x \approx \factRU\ x\ [x \ge \symb{1}]$ is an inductive theorem if
$(\factTU\ x)\gamma \approx (\factRU\ x)\gamma$ is convertible for those ground substitutions $\gamma$ that satisfy $\termInterpret{(x \ge \symb{1}) \gamma}=\true$. 
This includes, e.g., $[x:=\symb{1}]$ but not $[x:=\symb{0}]$. 

\myparagraph{Lemma generation}
A practical challenge in applying RI to prove concrete examples of program equivalence is the automatic generation of \emph{lemmas}. 
For example, when naively trying to prove $\factTU\ x \approx \factRD\ x\ [x \ge \symb{1}]$, we run successively into proof obligations
$\uploop\ x\ 3\ 2 \approx x * (\uploop\ x'\ 2\ 1)\ [x \geq 2 \wedge x' = x - 1]$, then
$\uploop\ x\ 4\ 6 \approx x * (\uploop\ x'\ 3\ 2)\ [x \geq 3 \wedge x' = x - 1]$, then
$\uploop\ x\ 5\ 24 \approx x * (\uploop\ x'\ 4\ 6)\ [x \geq 4 \wedge x' = x - 1]$, and so on.
This is called a \emph{divergence}.
Case analysis and rewriting brings us to the next equation in the sequence, but we can never use one as
an induction hypothesis to eliminate the next.  We need the insight that all these equations are instances
of, e.g., the lemma $\uploop\ x\ i\ a \approx x * (\uploop\ x'\ i'\ a')\ [x \geq i' \wedge x' = x - 1 \wedge
i = i' + 1 \wedge a = a' * i']$ to  obtain a proof goal matching the induction hypothesis, allowing the induction step to close the proof. 

This need for lemmas -- or to generalize equations before they can be used as induction hypotheses -- is a well-known topic in equational reasoning. 
Compared to equational reasoning, our situation is more involved, as the semantical reasoning over logical constraints (here integer expressions) requires discovering (loop) invariants. 
Two methods for constrained rewriting are provided by~\cite{fuh:kop:nis:17,hag:kop:23} and also apply to \lcstrss. 
For example, initialization generalization \cite{fuh:kop:nis:17} finds the lemma above, while matrix invariants \cite{hag:kop:23} allows us to prove $\factTU\ x \approx \factRU\ x\ [x \ge \symb{1}]$ by discovering the loop invariants $z = a * j$, $i=j+\symb{1}$ and $j \le x$ which are used to generate the lemma $\uploop\ x\ i\ z \approx a*(\symb{R}\ j\ x) \  [z = a * j \wedge i=j+\symb{1} \wedge j \le x]$. 

While these methods can successfully generate lemmas for many easy examples, they do not scale to more complicated computational behavior, which often require non-polynomial invariants. 
This leaves certain program equivalences beyond their scope. 
For example, the lemma generation methods in~\cite{fuh:kop:nis:17,hag:kop:23} are not able to prove  $\factRU\ x \approx \factTD\ x \ [x \ge \symb{1}]$,    
$\factRU\ x \approx \factRD\ x\  [x \ge \symb{1}]$ and $\factTU\ x \approx \factTD\ x$  
because the dependency between successive divergence steps cannot be expressed as a polynomial.
Fortunately, these equivalences can still be provable within RI. 
The key is not to focus on more expressive invariants, but rather to adopt a more high-level strategy: using templates we can generalize over loop structures themselves, instead of the specific values of loop variables, making these equivalences easily provable without complicated invariants.

\myparagraph{Templates}
Templates are 
rewrite schemes that capture commonly occurring programming patterns, such as upward or downward tail recursion. Templates can be modeled by higher-order recursors, which are rewrite systems.  Therefore, we can use RI to prove equivalences
between recursors, which can then be applied to prove concrete instances in practice. 
This provides a high-level lemma generation method, complementary to existing low-level methods from~\cite{fuh:kop:nis:17,hag:kop:23}. 
In this way, proving $\factTU\ x \approx \factTD\ x$ becomes equally easy as proving $\symb{sumTU}\ x \approx \symb{sumTD}\ x$, where this last equivalence involves a TailUp and TailDown implementation of $x \mapsto \sum_{i=1}^x i$. 
While the method in~\cite{hag:kop:23} can brute-force a lemma for $\symb{sumTU}\ x \approx \symb{sumTD}\ x$, this fails for $\factTU\ x \approx \factTD\ x$, as the required invariants are not polynomial. 
The template approach also allows us to prove other equivalences that were
previously out of reach. 

\subsection{Motivation}
\label{sec:motivation}
Templates in rewriting are not new. 
The most relevant related work -- and an inspiration for the present work -- 
is~\cite{chi:aot:toy:10}, which defines templates for verifying program transformations, including tail-recursive and non-tail-recursive implementations. However, this particular method suffers from several disadvantages that limit its applicability for analyzing real programming code, such as the examples in~\cref{fig:factcode}.
We now motivate that by using RI for \lcstrss, we can obtain a more practically useful template-based method for proving program equivalence. 
We focus on an illustrative example from~\cite{chi:aot:toy:10}, where a tail-recursion template is matched against a TRS computing the sum of a list of natural numbers:
\[
\begin{aligned}
    &
    \boxed{
    \begin{aligned}
    \mathsf{f}(u_4)
    &\to 
    \mathsf{f}_1(u_4, \mathsf{b})\\
    \mathsf{f}_1(\mathsf{a}, u_5)
    &\to 
    u_5\\
    \mathsf{f}_1(\mathsf{c}(u_6, v_6), w_6)
    &\to 
    \mathsf{f}_1(v_6, \symb{g}(w_6, u_6))
    \\
    \mathsf{g}(\symb{b}, u_7)
    &\to 
    u_7
    \\
    \mathsf{g}(\symb{d}(u_8, v_8), w_8)
    &\to 
    \mathsf{d}(u_8, \symb{g}(v_8, w_8))
    \end{aligned}}
    \qquad 
    \qquad 
    &
    \boxed{ 
    \begin{aligned}
    \mathsf{sum}(x_4)
    &\to 
    \mathsf{sum1}(x_4, \mathsf{0})\\
    \mathsf{sum1}([], x_5)
    &\to 
    x_5\\
    \mathsf{sum1}(x_6:y_6, z_6)
    &\to 
    \mathsf{sum1}(y_6, +(z_6, x_6))
    \\
    +(\symb{0}, x_7)
    &\to 
    x_7
    \\ 
    +(\symb{s}(y_8), z_8)
    &\to 
    \mathsf{s}(+(y_8, z_8))
    \end{aligned}}\\
    &\textit{Template (Tail Recursion)}
    &\textit{Matchable example (summing a list)}
\end{aligned}
\]

\myparagraph{Constrained rewriting}
The template is defined in (first-order) unconstrained term
rewriting.  This suffices for reasoning about programs on \emph{inductive} data structures through pattern matching
(like natural numbers and lists), but less so for structures like 
integers where cases are often investigated through constraints rather than pattern matching.
Hence, the tail recursion template does not instantiate to $\factTU\ x$
(or $\factTD\ x$), as there is no obvious correspondence between the
constructor-based recursion in $\symb{f}_1$ and the constraint-based
recursion of $\uploop$. 

\myparagraph{RI integration \& recursors}
We also notice that the template effectively represents a complete program description. 
We can subdivide the template in three parts:
\begin{enumerate}[label = (\arabic*).]
\item \emph{Initialization}: the rule 
$\mathsf{f}(u_4)
\to 
\mathsf{f}_1(u_4, \mathsf{b})$ initializes the accumulator $\symb{b}$
\item \emph{Recursive part}:  the two $\symb{f}_1$-rules 
\item \emph{Auxiliary function definition}: the two $\symb{g}$-rules
\end{enumerate}  
As we will see, integrating templates into RI allows us to focus solely on the recursive subsystem defined by $\symb{f}_1$, which we refer to as a \emph{recursor}.
Reasoning about initialization and auxiliary functions can be delegated to the RI system.
In this way, we obtain a clearly focused template-based proof method centered on recursors.

\myparagraph{Broader applicability via higher-order rewriting}
The restriction to \emph{first-order} term rewriting limits its applicability to only a subset of matchable instances. For example, in the above template the function
symbol $\symb{g}$ may capture symbols like $+$  or $\symb{append}$, but
does not capture multiplication.  
With higher-order rewriting, we can 
represent $\symb{f}_1$ as an instance of the higher-order 
$\symb{foldl}$ recursor, defined as 
\begin{center}
\(
\symb{foldl}\ f\ a\ [] 
\to 
a
\qquad \qquad 
\symb{foldl}\ f\ a\ (x:xs) 
\to 
\symb{foldl}\ f\ (f\ a\ x)\ xs 
\)
\end{center}
This is useful because
\begin{itemize}
\item 
$\funcvar :: \int \to \int \to \int$
  is treated as a higher-order parameter to $\symb{foldl}$. Therefore, it is defined for \emph{arbitrary} instances of $f$. 
\item Specific instantiations of auxiliary functions such as $f:=+, *$ are handled during the RI procedure. 
\end{itemize}
That is, we can take advantage of the higher-order nature of \lcstrss\ to
reuse the same recursors with entirely different instantiations.
In the above example, we can capture  
${\symb{sum1}\ l\ a} \approx {\symb{foldl}\ \prefix{+}\ a\ l}$ as a
template-recursor equivalence.  
The same idea applies to list recursors: we can use $\symb{foldl}$ (and its commonly-used
counterpart $\symb{foldr}$) as a recursor to represent a tail recursion
template on lists.

\subsection{Contributions}
As motivated above, the specific combination of templates with \lcstrss\ opens up possibilities for integration in the formal RI system in a manner very useful for real program verification: 
\begin{itemize}[label = $\triangleright$] 
\item  
Since \lcstrs\ are
a form of higher-order rewriting, we can perform all equivalence proofs
of recursors within RI, rather than at the meta-level.
This enables future integration of the template method into rewrite tools as a lemma generation technique complementing existing approaches for generalizing lower-level behavior.
For now, all template-based proofs in this paper are manually verified in \cora. 
This is a termination tool for \lcstrss\ and also implements Bounded RI~\cite{hag:kop:26}. 
\item Since constrained rewriting directly supports primitive data structures, the rewrite rules closely resemble real code. 
This in contrast to the required encodings in unconstrained rewriting, which would make analysis of loop constructions (like in our factorial implementations)
infeasible or at least inconvenient. 
With \lcstrss\ we can define suitable templates in a human-readable manner. 
\end{itemize}

\myparagraph{Overview}
After we recap \lcstrss\ and Bounded RI in~\Cref{sec:prelim}, we continue as follows 
\begin{itemize}[label = $\triangleright$]
\item In~\Cref{sec:templates} we introduce four templates and describe how to match them with concrete programs. 
\item In \Cref{sec:temp-eq} we assume recursor equivalences as given and focus on the proof-tactical aspect of how to apply these equivalences to prove concrete template instances.
We distinguish two proof tactics: \emph{one-sided matching}, transforming a proof goal into an easier one, and \emph{two-sided matching}, concluding equivalence directly.
This section is meant to provide an intuitive explanation.
\item In \Cref{sec:temp-RI-integration} we explain the technical requirements needed to use templates in Bounded RI as a lemma generation method.
We discuss the corresponding ordering requirements that constitute the separate parts of template-based proofs, and describe how to recombine them in concrete instances. 
\end{itemize}

\section{Preliminaries}
\label{sec:prelim}
\lcstrss~\cite{guo:kop:24} are a higher-order rewriting formalism with built-in support for theories
such as integers and boolean (though in fact, any arbitrary theory such as bitvectors, floating
point numbers or integer arrays is in principle supported) as well as logical constraints to model
control flow. 
This considers \emph{applicative} higher-order term rewriting 
(without $\lambda$ abstractions) 
and \emph{first-order} constraints. 

\myparagraph{Types}
Assume given a set of sorts (base types) $\Sorts$. 
The set $\Types$ of types is defined by
\(
    \Types
    ::=
    \Sorts
    \mid
    \Types \arrtype \Types
\).
Here, $\arrtype$ is right-associative: all types may be written as
$\mathit{type}_1 \arrtype \dots \arrtype \mathit{type}_m \arrtype \mathit{sort}$
with $m \geq 0$.
We also assume given a subset $\thSorts \subseteq \Sorts$ of \emph{theory sorts} with
$\bool \in \thSorts$, and define the \emph{theory types} by
\(
    \thTypes
    ::=
    \thSorts
    \mid
    \thSorts \to \thTypes
\).
Every $\iota \in \thSorts$ has an interpretation set $\typeInterpret{\iota} \ne \emptyset$.  

In this paper, we fix $\thSorts \supseteq \{\int,\bool\}$, with $\typeInterpret{\bool} =
\{\true,\false\}$ and $\typeInterpret{\int} = \mathbb{Z}$ (the set of all integers). 
For any sort $\asort$ and type $\atype$, we let $\typeInterpret{\asort \arrtype \atype}$ be the set
of all total functions from $\typeInterpret{\asort}$ to $\typeInterpret{\atype}$.

\myparagraph{Terms}
We assume given a signature $\Sig$ of \emph{function symbols} and a disjoint set $\Var$ of
variables, and a function $\typeof$ from $\Sig \cup \Var$ to $\Types$.
The set of terms $\Terms$ over $\Sig$ and $\Var$ are those expressions which can be typed using 
the recursive clauses:
(a) $u :: \typeof(u)$ for $u \in \Sig \cup \Var$; and
(b) $s\ t :: \btype$ of $s :: \atype \arrtype \btype$ and $t :: \atype$.
Application is left-associative; every term may be written as
$u\ s_1 \cdots s_m$ with $u \in \Sig \cup \Var$ and $m \geq 0$.
For a term $t$, let $\Vars{t}$ be the set of variables in $t$.
A term $t$ is \emph{ground} if $\Vars{t} = \emptyset$.
We assume that $\Sig$ is the disjoint union $\thSig \uplus \termsSig$, where
$\typeof(\afun) \in \thTypes$ for all $\afun \in \thSig$.

Each $\afun \in \thSig$ has an interpretation $\termInterpret{\afun} \in \typeInterpret{\typeof(\afun)}$.
Terms in $T(\thSig, \Var)$ are \emph{theory terms}.
For \emph{ground} theory terms, we define
$\termInterpret{s\ t} = \termInterpret{s}(\termInterpret{t})$, thus mapping each ground term of type
$\atype$ to an element of $\typeInterpret{\atype}$.
For example, we have theory terms 
$x + \symb{3}$, $\strue$ and $\symb{7} * \symb{0}$.
The latter two are ground, and we have $\termInterpret{\strue} = \true$ and  
$\termInterpret{\symb{7} * \symb{0}} = 0$.
\emph{Values} are theory symbols of base type, i.e.
\(
    \Val
    =
    \{
        v \in \thSig
        \mid
        \typeof(v) \in \thSorts
    \}
\).
A \emph{constraint} is a theory term $\varphi :: \bool$, such that $\typeof(\avar) \in \thSorts$
for all $x \in \Vars{\varphi}$.

In this paper, we fix a theory signature $\thSig \supseteq
\{+,-,*,=,<,\le,>,\geq,\wedge,\vee,\neg,\strue,\sfalse\} \cup \{\symb{n} \mid n \in \mathbb{Z}\}$
with each of these symbols interpreted as expected, e.g. 
$*::\int \to \int \to \int
$ is interpreted as
multiplication on $\mathbb{Z}$.
We use $\prefix{\afun}$ for prefix or partially
applied notation, e.g. $\prefix{+}\ x\ y$ and $x + y$ are the same.
The values in this setting are (at least) $\strue,\sfalse$ and all $\mathsf{n}$.
An example of a constraint is $x*y>\symb{0}$. 

\myparagraph{Contexts and context functions}
Let $\square_1,\dots,\square_n$ be fresh, typed constants, with $n \geq 1$.
A context $C[\square_1,\dots,\square_n]$ (or just: $C$) is a term in $T(\Sig \cup \{\square_1, \ldots, \square_n\}, \Var)$
in which each $\square_i$ occurs exactly once.
The term obtained from $C$ by replacing each $\square_i$ by a term $t_i$ of the same type is denoted by $C[t_1,\dots,t_n]$.

A \emph{context function} in $n$ parameters is a term $C(\square_1,\dots,
\square_n)$ (or just: $C$) in $T(\Sig \cup \{\square_1, \ldots, \square_n\},
\Var)$ in which each $\square_i$ may occur 0 or more times.  The term
obtained from $C$ by replacing each $\square_i$ by a term $t_i$ of the same
type (but multiple instances of the same $\square_i$ are replaced by the
same $t_i$) is denoted $C(t_1,\dots,t_n)$.

A context function in $2$ parameters is called a \emph{binary} context
function.

\myparagraph{Rewrite rules}
A rule is an expression $\myrule$.
Here $\ell$ and $r$ are terms of the same type, $\ell$ has a form $\afun\ \ell_1 \cdots \ell_k$ with
$\afun \in \Sig$ and $k \geq 0$, $\varphi$ is a constraint and
\( \Vars{r} \subseteq \Vars{\ell} \cup \Vars{\varphi} \).
If $\varphi = \strue$, we may denote the rule as just $\ell \to r$.
Fixing a signature $\Sig$, we assume given a set of rules $\rules$ whose
left-hand sides are not theory terms, and define the set of
\emph{calculation rules} as:
\(
\calcrules 
=
\{
\afun\ x_1 \cdots x_m \to y \ [ y = \afun\ x_1 \cdots x_m ] 
\ \mid\ 
\afun \in \thSig \setminus \Val,\ 
x_1, \ldots, x_m \in \Var
,
y \in \Var,\ \afun :: \asort_1 \arrtype \dots \arrtype \asort_m \arrtype \bsort
\}
\).
We require that every symbol $\afun$ has a unique \emph{arity}
$\arity(\afun)$: if there is a rule $\afun\ \ell_1 \cdots \ell_k \arrz
r\ [\varphi]$ in $\rules \cup \calcrules$ then $\arity(\afun) = k$, and
if there is no such rule then $\arity(\afun) = \infty$.  A function symbol
is a \emph{constructor} if $\arity(\afun) = \infty$.
A term is a \emph{semi-constructor term} if, for all its subterms
$\afun\ s_1\ \cdots s_n$, we have $n < \arity(\afun)$.
(Hence, a semi-constructor term is irreducible.)

\myparagraph{Reduction relation}
We assume familiarity with substitutions and denote $\domain(\gamma) = \{x \in \Var \mid \gamma(x) \ne x\}$. 
A substitution \( \gamma \) \emph{respects} a constraint \( \varphi \) if
$\gamma(x) \in  \Val$ for all $x \in \Vars{\varphi}$ and $\termInterpret{\varphi\gamma} = \true$.

For a signature $\Sig$ and set of rewrite rules $\rules$, the reduction relation
$\rw$ is defined by:
\[
C[l \gamma]
\rw
C[r \gamma]
\text{ if }
\myrule \in
\rules \cup \calcrules
\text{ and }
\gamma
\text{ respects }
\varphi
\]
For example, in \Cref{fig:RD} we have  
$
\factRD\ \symb{2} 
\rw 
\symb{2}*(\factRD\ (\symb{2}-\symb{1})) \rw 
\symb{2}*(\factRD\ \symb{1})
\rw
\symb{2}*\symb{1}
\rw 
\symb{2}
$. 

\myparagraph{Inductive theorems}
An \emph{equation} is a triple $s \approx t \ [\varphi]$ with
$\typeof(s) = \typeof(t)$ 
and $\varphi$ a constraint. 
A substitution $\gamma$ respects $s \approx t \ [\varphi]$ if $\gamma$ respects $\varphi$ 
and $\Vars{s} \cup \Vars{t} \subseteq \domain(\gamma)$.
A \emph{ground substitution} is a substitution \( \gamma \) such that \( \gamma(x) \) is a ground term, for all
\( x  \in \domain(\gamma)\).
 
An equation $s \approx t \ [\varphi]$ is an \emph{inductive theorem} (or \emph{ground convertible}) 
if $\rweq{s \gamma}{t \gamma}{\rules}$ for every ground substitution $\gamma$ that respects it.
Here $\leftrightarrow_\rules \ =\  \rightarrow_\rules \cup \leftarrow_\rules$,
and $\leftrightarrow_\rules^*$ is its transitive, reflexive closure. 

\myparagraph{Rewriting Induction}
Rewriting Induction is a proof system to show that all elements of a set $\eqs$ of equations are
inductive theorems.  There are several variations of the core proof system, but typically, they
consist of a set of derivation rules that allow the user to iteratively transform a pair
$(\eqs,\hs)$ of \emph{equations} and \emph{induction hypotheses} until all equations are eliminated.
For example, derivation steps might include rewriting an equation using a rule or induction
hypothesis, doing case analysis on some variable, or deleting a vacuous equation $s \approx s\ 
[\varphi]$.  Variations of RI may  differ along several aspects: the exact derivation rules, whether they support constraints~\cite{fal:kap:12,fuh:kop:nis:17, hag:kop:24, hag:kop:26} or not~\cite{aot:06, aot:08a, aot:08b, aot:toy:16, red:90}, whether they 
transform a triple $(\eqs,\hs,\mathcal{G})$ instead of a pair~\cite{aot:06, aot:08a},
or what exactly $\eqs$ and $\hs$ contain (e.g., in \cite{hag:kop:26} the equations in $\eqs$ carry additional information and the elements of $\hs$ are equations rather than rules).

All such systems operate under the assumptions of \emph{termination} and
\emph{quasi-reductivity}: the relation $\rw$ may not admit infinite
reductions (hence, in the setting of RI we only use \lcstrss\ to model terminating programs), and ground terms may only be irreducible if they are
semi-constructor terms (essentially: there are no missing cases in the
pattern matching).
Moreover, a central component in these systems is the incorporation of
well-founded induction: in essence,
the system proves the given theorems by a shared induction on a well-founded ordering $\succ$.
However, the specific construction of such an ordering $\succ$ is realized differently across existing RI variants.
In many cases (e.g., \cite{fal:kap:12,fuh:kop:nis:17,hag:kop:24,red:90}) the ordering is
implicitly defined as $\rw \cup \to_{\hs}$, where $\hs$ consists of
\emph{rules} (which requires termination of $\to_{\rules \cup \hs}$).
In others (e.g., \cite{aot:06, aot:08a, aot:08b, aot:toy:16}) the user is
required to explicitly provide an ordering, which is then used in certain
derivation steps; typically, this ordering should include $\rw$.

\myparagraph{Bounded Rewriting Induction}
The template method introduced in this paper is formulated so as to be compatible with implementation in \emph{Bounded RI}~\cite{hag:kop:26} for \lcstrss.
This system follows an approach similar to those in \cite{aot:06, aot:08a, aot:08b, aot:toy:16}, and aims to minimize the number of ordering requirements, thereby increasing the practical applicability of the template method. 
Bounded RI is implemented in the rewriting tool \cora, which enables the formal verification of template-based equivalence proofs and facilitates future automation.
Although we cannot present the full Bounded RI system in this paper, we nevertheless discuss the technical  conditions that must be satisfied: using \emph{bounded ground convertibility} we can present the template method in a self-contained manner (independent of the specific language of Bounded RI).
To define this notion, we first assume a fixed well-founded ordering $\succ$ that includes $\rw \cup
\supterm$, where $\supterm$ is the subterm relation, generated
by $s\ t \supterm s$ and $s\ t \supterm t$.\footnote{This is
  not strictly required for the application of Bounded RI~\cite{hag:kop:26}. 
  It suffices if $\rw \cup \supterm$ is included in a corresponding
  quasi-ordering $\succeq$, and $\rw$ does not need to be terminating.
  We simplify the requirements here for ease of presentation.}
Let $\succeq$ be the reflexive closure of $\succ$.
Let $s \succ t\ [\varphi]$ denote that $s\gamma \succ t\gamma$ for all ground
substitutions $\gamma$ that respect $\varphi$ and satisfy $\Vars{s} \cup \Vars{t}
\subseteq \domain(\gamma)$.  
Define $s \succeq t\ [\varphi]$ if either $s = t$ or
$s \succ t\ [\varphi]$.

\myparagraph{Bounded ground convertibility}
Bounded RI is designed to prove \emph{bounded} ground convertibility which implies ground convertibility and has additional applications in
proving ground confluence~\cite{hag:kop:26}.

For terms $\sterm,\tterm$ 
we say that $s \approx t\ [\varphi]$ is \emph{$\rules$-bounded ground convertible} under $\sterm,\tterm$, denoted
$\rconv{s}{t}{\sterm,\tterm}\ [\varphi]$, if $s=t$ or for every
ground substitution $\gamma$ that respects $\varphi$ there exist ground
contexts $C_0,\ldots,C_n$ and ground terms
$s_0,\ldots,s_n,t_0,t_1,\ldots,t_n$ so that $s\gamma = C_0[s_0]$,
$t\gamma = C_n[t_n]$, and for all $0 \leq i \leq n$:
(a) $s_i \leftrightarrow_{\rules} t_i$,
(b) $\sterm\gamma \succeq s_i$ or $\tterm\gamma \succeq s_i$,
(c) $\sterm\gamma \succeq t_i$ or $\tterm\gamma \succeq t_i$,
(d) if $i > 0$: $C_{i-1}[t_{i-1}] = C_i[s_i]$.
(So
$s\gamma = C_0[s_0] \leftrightarrow_{\rules} C_0[t_0] = C_1[s_1] \leftrightarrow_{\rules} \ldots
\leftrightarrow_{\rules} C_n[t_n] = t\gamma$,
and each $s_i,t_i$ is dominated by at least one of $\sterm\gamma$ or
$\tterm\gamma$.)

We say $s \approx t\ [\varphi]$ is a \emph{bounded inductive theorem}
if $\rconv{s}{t}{s,t}\ [\varphi]$.

\section{Templates}
\label{sec:templates}
A template is a rewrite scheme that captures a particular programming construct.
\Cref{tab:templates} presents four templates that generalize the constructs used in the factorial implementations, describing recursion and tail recursion on integers in either upward or downward direction.
The templates are defined by parametrized contexts, whose parameters include a context function $\ff{\square_1}{\square_2} :: \int$ (used in the recursive call), and terms $\lowterm :: \int$ or $\upterm :: \int$ (used as lower or upper bound), as well as possibly a term $\baseterm::\int$ (used as a base case).
The symbols $T$ and $R$ indicate tail-recursion and general recursion, while the superscripts $+$ and $-$ indicate upward and downward direction, respectively. 
To be exact:

$\tutemp[\square_1,\square_2], \tdtemp[\square_1,\square_2] :: \int$ are contexts with $\square_1, \square_2 :: \int$.

$\rutemp[\square], \rdtemp[\square] :: \int$ are contexts with $\square :: \int$.

The parameter $\ff{\square_1}{\square_2} :: \int$ is a context function with $\square_1,\square_2 :: \int$, while
$\lowterm,\upterm,\baseterm :: \int$ are terms.

Moreover, the template structure employs one or two integer \emph{variables}: $\ivar :: \int$ and $\accvar :: \int$.  These variables may not occur inside the other terms and contexts.

\begin{center}
\begin{tabular}{l|l|l}
\textbf{Construct} 
& 
\textbf{Template} 
&
\textbf{Computed term}\\
\noalign{\hrule height 1pt}
TailUp
&
$
\begin{aligned}
\tutemp[\ivar, \accvar] & \to \accvar & [\ivar>\upterm] \\
\tutemp[\ivar, \accvar] & \to \tutemp[\ivar+\symb{1}, \ff{\ivar}{\accvar}] & [\ivar \le \upterm]
\end{aligned}
$
&
$
\ff{\upterm}{\ff{\upterm - 1}{
\cdots ,
\ff{\ivar+1}{\ff{\ivar}{\accvar}}
\cdots
}}
$
\\
\hline 
TailDown
&
$
\begin{aligned}
\tdtemp[\ivar, \accvar] & \to \accvar & [\ivar<\lowterm] \\
\tdtemp[\ivar, \accvar] & \to \tdtemp[\ivar-\symb{1}, \ff{\accvar}{\ivar}] & [\ivar \ge \lowterm]
\end{aligned}
$
&
$
\ff{\ff{
        \cdots 
        \ff{\ff{\accvar}{\ivar}}{\ivar-1}
        \cdots  
}{\lowterm+1}}{\lowterm}
$
\\
\noalign{\hrule height 1pt}
RecUp
&
$
\begin{aligned}
\rutemp[\ivar] & \to \baseterm & [\ivar>\upterm] \\
\rutemp[\ivar] & \to \ff{\rutemp[\ivar+\symb{1}]}{\ivar} & [\ivar \le \upterm]
\end{aligned}
$
&
$
\ff{\ff{\cdots\ff{\ff{\baseterm}{\upterm}}{\upterm-1} \cdots}{\ivar+1}}{\ \ivar}
$
\\
\hline
RecDown
&
$
\begin{aligned}
\rdtemp[\ivar] & \to \baseterm & [\ivar<\lowterm] \\
\rdtemp[\ivar] & \to \ff{\ivar}{\rdtemp[\ivar-\symb{1}]} & [\ivar \ge \lowterm]
\end{aligned}
$
&
$\ff{\ivar}{\ff{\ivar-1}{\cdots, \ff{\lowterm+1}{\ff{\lowterm}{\baseterm}} \cdots}}$
\\
\noalign{\hrule height 1pt}
\end{tabular}
\captionof{table}{Templates for upward/downward recursion and tail recursion}
\label{tab:templates}
\end{center}

Here,
$\ivar$ serves as a loop index, which is decreased or increased by
$1$ at each recursive call, until $i$ is below the lower bound $\lowterm$ or above the upper bound $\upterm$.
At that point, the computation terminates and returns either the base case $\baseterm$, or the variable $\accvar$ which serves as an \emph{accumulator}.
The last column of \Cref{tab:templates} displays a schematic presentation of the term being computed during this procedure. 

\begin{example}\label{ex:template-matching}
We recall the TU, TD, RU, RD implementations of the factorial function from \cref{fig:factlcstrs}.
We demonstrate how to match them with the corresponding templates.

\Cref{fig:TU} is straightforward: we choose $\ff{\square_1}{\square_2} := \square_1 * \square_2$, $\upterm := x$ and $T^+_{\funccont,x}[\square_1,\square_2] := \uploop\ x\ \square_1\ \square_2$.

Similarly \cref{fig:RU}: we choose $\ff{\square_1}{\square_2} := \square_2 * \square_1$, $\upterm := x - 1$, $\baseterm := x$ and $R^+_{F,x-1,x}[\square] := \uprec\ \square\ x$.

In these cases, the template cases exactly capture the rules for
$\uploop$ and $\uprec$ respectively.  However, for TD (\cref{fig:TD}) and RD (\cref{fig:RD}), we do not have an exact syntactic match with the template shapes in \Cref{tab:templates} due to the inequality-mismatch in the constraints.
However, since we are dealing with integers, this is solved by observing that $\ivar \leq n$ is equivalent to $\ivar < n+1$ and $\ivar > n$ equivalent to $\ivar \geq n+1$.  Hence, we can pre-process these two \lcstrss\ to obtain equivalent systems:

\begin{minipage}[t]{0.45\textwidth}
\(\boxed{
\begin{aligned} 
&
\factTD\ x
\to 
\downloop\ x\ \symb{1}
\\
&
\downloop\ x\ a
\to 
a
&
[x < \symb{1}]
\\
&
\downloop\ x\ a
\to 
\downloop\ (x-\symb{1})\ (a*x)
&
[x \ge \symb{1}]
\end{aligned}}
\)
\end{minipage}
\begin{minipage}[t]{0.45\textwidth}
\(\boxed{
\begin{aligned}
\factRD\ x
&
\to 
\symb{1}
&
[x < \symb{2}]
\\
\factRD\ x
&
\to 
x*(\factRD\ (x-\symb{1}))
&
[x \ge \symb{2}]
\end{aligned}}
\)
\end{minipage}

In this representation, we easily match with the templates
$T^-_{\square_1 * \square_2,\symb{1}}$ and
$R^-_{\square_1 * \square_2,\symb{2},\symb{1}}$ respectively, using the
variable $x$ for the index $i$.
\end{example}

\subsection{Recursors}\label{subsec:recursors}
To generalize the template shapes, we introduce four corresponding 
higher-order recursors: $\rulesrec=$

\noindent
\[
\left\{
\begin{array}{lrclllrcll}
\tub & \tailup\ \funcvar\ \ivar\ \upvar\ \accvar & \to & \accvar &
  [\ivar > \upvar] &
\tur & \tailup\ \funcvar\ \ivar\ \upvar\ \accvar & \to &
  \tailup\ \funcvar\ (\ivar+\symb{1})\ \upvar\ (\funcvar\ \ivar\ \accvar)
  & [\ivar \leq \upvar] \\
\tdb & \taildown\ \funcvar\ \lowvar\ \ivar\ \accvar & \to & \accvar &
  [\ivar < \lowvar] &
\tdr & \taildown\ \funcvar\ \lowvar\ \ivar\ \accvar & \to &
  \taildown\ \funcvar\ \lowvar\ (\ivar - \symb{1})\ (\funcvar\ \accvar\ 
  \ivar) & [\ivar \geq \lowvar] \\
\rub & \recup\ \funcvar\ \ivar \ \upvar\  \basevar & \to & \basevar &
  [\ivar > \upvar] &
\rur & \recup\ \funcvar\ \ivar \ \upvar\ \basevar & \to &
  \funcvar\ (\recup\ \funcvar \ (\ivar+\symb{1})\ \upvar\ \basevar)\ 
  \ivar & [\ivar \leq \upvar] \\
\rdb & \recdown\ \funcvar\ \lowvar\ \ivar\ \basevar\ & \to & \basevar &
  [\ivar < \lowvar] &
\rdr & \recdown\ \funcvar\ \lowvar\ \ivar\ \basevar & \to &
  \funcvar\ \ivar\ (\recdown\ \funcvar\ \lowvar\ (\ivar -
  \symb{1})\ \basevar) & [\ivar \geq \lowvar] \\
\end{array}
\right\}
\]

With Bounded RI, it is trivially easy to prove for instance that
$\uploop\ \upvar\ \ivar\ \accvar \approx \tailup\ \prefix{*}\ \ivar\ \upvar\ 
\accvar\ [\strue]$ is an inductive theorem.
More generally, we have the following result:

\begin{lemma}[Template-recursor equivalences]\label{lem:temp-rec-eq}
Consider a quasi-reductive \lcstrs\ with rules $\rules$, and a well-founded
ordering $\succ$ that includes $\rw \cup \supterm$.  Assume given a context
function $\funccont$ and terms $\upterm,\lowterm,\baseterm$ (in which the
variables $\ivar$, $\accvar$ do not occur), and suppose that $\rules$
contains a rule $\Qsymb\ \avar\ \bvar \to \ff{\avar}{\bvar}$.
Then:
\begin{itemize}[label = $\triangleright$]
\itemsep0em
\item if $\rules$ includes $\tub$, $\tur$ and the rules defining TailUp for
  some context $\tutemp$, \\
  then $\tutemp[\ivar,\accvar] \approx
  \tailup\ \Qsymb\ \ivar\ \upterm\ \accvar\ [\strue]$
  is a bounded inductive theorem;
\item if $\rules$ includes $\tdb$, $\tdr$ and the rules defining TailDown for
  some context $\tdtemp$, \\
  then $\tdtemp[\ivar,\accvar] \approx
  \taildown\ \Qsymb\ \lowterm\ \ivar\ \accvar\ [\strue]$
  is a bounded inductive theorem;
\item if $\rules$ includes $\rub$, $\rur$ and the rules defining RecUp for
  some context $\rutemp$, \\
  then $\rutemp[\ivar] \approx
  \recup\ \Qsymb\ \ivar\ \upterm\ \baseterm\ [\strue]$
  is a bounded inductive theorem;
\item if $\rules$ includes $\rdb$, $\rdr$ and the rules defining RecDown for
  some context $\rdtemp$, \\
  then $\rdtemp[\ivar] \approx
  \recdown\ \Qsymb\ \lowterm\ \ivar\ \baseterm\ [\strue]$
  is a bounded inductive theorem.
\end{itemize}
Moreover, if $\ff{\square_1}{\square_2}$ is just $\afun\ \square_1\ 
\square_2$ for some $\afun$, then all above statements hold with $\afun$
in place of $\Qsymb$.
\end{lemma}

\myparagraph{RI integration}
The equivalences from \lemref{temp-rec-eq} describe the formal connection between templates and their recursor representation. 
This is important for the Bounded RI integration: identifying a concrete program as a template instance amounts to identifying it as a corresponding recursor instance. Since recursors are described by \lcstrss, this reduces to a matching problem, which is a syntactical problem.
We do not foresee any problems in implementing such a matching algorithm.

\section{Applying Recursor Equivalences}
\label{sec:temp-eq}
Having matched each of the factorial implementations with a corresponding template in \Cref{tab:templates}, and thereby with the higher-order recursors
of \Cref{subsec:recursors}, two major challenges remain: deriving equivalences between the various recursors, and applying them in practice to prove program equivalence. 
In this section, we focus on the latter from an informal and proof-strategic perspective.
In \Cref{sec:temp-RI-integration} we will
discuss how to formalize this informal reasoning within Bounded RI.

\myparagraph{Two easy recursor equivalences}
Two recursor equivalences are immediately suggested by the computational structure of the templates:
\begin{itemize}[label = $\triangleright$]
\item  
TailUp and RecDown: 
the current index $\ivar$
appears as the first argument of the function context $\funccont$, with the
accumulator or recursive call being the second at each step.
In particular, $\tutemp[\lowterm, \baseterm]$ and $\rdtemp[\upterm]$ both compute 
$\ff{\upterm}{\ff{\upterm - 1}{
\cdots ,
\ff{\lowterm+1}{\ff{\lowterm}{\baseterm}}
\cdots
}}$
\item 
TailDown and RecUp: 
each recursive step introduces a new outer application of $\funccont$. 
In particular, $\tdtemp[\upterm,\baseterm]$ and $\rutemp[\lowterm]$ both
compute
$
\ff{\ff{\cdots\ff{\ff{\baseterm}{\upterm}}{\upterm-1} \cdots}{\lowterm+1}}{\ \lowterm}
$. 
\end{itemize}
\begin{lemma}\label{lem:uncond-temp-eq}
Consider a quasi-reductive \lcstrs\ with $\rules \supseteq
\rulesrec$, and a well-founded ordering
$\succ\;\supseteq\;(\rw \cup \supterm)$ such that
$\tailup\ \funcvar\ \lowvar\ \upvar\ \cvar \succ \recdown\ \funcvar\ \lowvar'\ 
\upvar\ (\funcvar\ \lowvar\ \cvar)\ [\lowvar \leq \upvar \wedge \lowvar' = \lowvar + \symb{1}]$
and
$\taildown\ \funcvar\ \lowvar\ \upvar\ \cvar \succ \recup\ \funcvar\ \lowvar\ \upvar'\ 
(\funcvar\ \cvar\ \upvar)\ [\lowvar \leq \upvar \wedge \upvar' = \upvar - \symb{1}]$.
Then the following equations are bounded inductive theorems:
\[
\begin{aligned}
\TURDlem
\ \ 
\tailup\ \funcvar\ \lowvar\ \upvar\ \cvar
\approx 
\recdown\ \funcvar\ \lowvar\ \upvar\ \cvar
\qquad 
\text{ and }
\qquad 
\TDRUlem
\ \  
\taildown\ \funcvar\ \lowvar\ \upvar\ \cvar
\approx
\recup\ \funcvar\ \lowvar\ \upvar\ \cvar
\end{aligned}
\]
\end{lemma}

We now illustrate how to apply \lemref{uncond-temp-eq} to prove $\factTU\ x \approx \factRD\ x$ and $\factTD\ x \approx \factRU\ x \ [x \ge \symb{1}]$.
We can distinguish two approaches: \emph{one-sided matching} or \emph{two-sided matching}. 

\subsection{One-sided matching}
\label{sec:onesided-matching}

In one-sided matching, we apply both a template-recursor equivalence and a
recursor equivalence to reduce one side of an equation to obtain an easier
proof goal.
Depending on the side of the equation and the recursor equivalence applied, the resulting proof goals may vary in difficulty.

\begin{example}
\label{ex:onesided-matching1}
Consider the equation $\factTU\ x \approx \factRD\ x\ [x \geq \symb{1}]$.
We can rewrite the left-hand side in 4 steps, obtaining the new equation
$\uploop\ x\ \symb{2}\ \symb{1} \approx \factRD\ x\ [x \geq \symb{1}]$.%
We may then apply one-sided matching with $\TURDlem
\ 
\tailup\ \funcvar\ \lowvar\ \upvar\ \cvar
\approx 
\recdown\ \funcvar\ \lowvar\ \upvar\ \cvar$ from 
\lemref{uncond-temp-eq} to either change the left side or the right side:
\begin{itemize}[label=$\triangleright$]
\item Left side: in \exaref{template-matching} we established \( \uploop\ x\ \symb{2}\ \symb{1} =
    T^+_{\square_1 * \square_2,x}[\symb{2},\symb{1}]
  \). By \lemref{temp-rec-eq},
  \( T^+_{\square_1 * \square_2,x}[\symb{2},\symb{1}] \approx
    \tailup\ \prefix{*}\ \symb{2}\ x\ \symb{1} \) is an inductive theorem.
 So it suffices to show \( \tailup\ \prefix{*}\ \symb{2}\ x\ \symb{1} \approx
  \factRD\ x\ [x \geq \symb{1}] \).
  Applying $\TURDlem$ to the left side of this last equation, we obtain 
  \( \eq{E}{1}\ \recdown\ \prefix{*}\ \symb{2}\ x\ \symb{1} \approx
  \factRD\ x\ [x \geq \symb{1}] \).
\item Right side: similarly, we write 
  \( \factRD\ x = R^-_{\square_1 * \square_2,\symb{2},\symb{1}}\ x \) and
  have an inductive theorem $R^-_{\square_1 * \square_2,\symb{2},
  \symb{1}}\ x \approx \recdown\ \prefix{*}\ \symb{2}\ x\ \symb{1}\ [x \geq
  \symb{1}]$. 
  Therefore, we should prove $\uploop\ x\ \symb{2}\ \symb{1} \approx \recdown\ \prefix{*}\ \symb{2}\ x\ \symb{1}\ [x \geq \symb{1}]$. 
  Applying $\TURDlem$ to the right-hand side of this last equation, we obtain
  \(\eq{E}{2} \ \uploop\ x\ \symb{2}\ \symb{1} \approx \tailup\ \prefix{*}\ \symb{2}\ x\ 
  \symb{1} \).
\end{itemize}
\end{example}

In both cases in \exaref{onesided-matching1} we applied a recursor
equivalence to produce an equation that compares two terms of similar
iterative structure (e.g. both upward tail recursive). 
This was effective because
% , 
% using low-level techniques, 
such comparisons are often much simpler to prove than comparisons between different programming
structures: both $\eq{E}{1}$ and $\eq{E}{2}$ can be proved in an
entirely straightforward way (without requiring further lemma equations or
generalizations).
Such proofs are typically easy to be done in a fully automatic fashion (although
we have not yet implemented this automation in \cora).
Unfortunately, the strategy does not always suffice on its own, as
illustrated in \exaref{onesided-matching2}.

\begin{example}\label{ex:onesided-matching2}
To prove the inductive theorem \( \factTD\ x \approx \factRU\ x\ [x \geq \symb{1}] \), existing lemma generation methods for constrained
rewriting (e.g.,\cite{fuh:kop:nis:17,hag:kop:23}) do not suffice.
We try making progress with one-sided matching.
First, we reduce both sides of the equation, obtaining
\( \downloop\ (x - \symb{1})\ x \approx \uprec\ \symb{1}\ x\ [x \geq
\symb{1}] \).  Then:
\begin{itemize}[label=$\triangleright$]
\item Left side: using the same strategy as in \exaref{onesided-matching1}, we obtain the proof
  goal \( \recup\ \prefix{*}\ \symb{1}\ (x - \symb{1})\ x \approx
  \uprec\ \symb{1}\ x\ [x \geq \symb{1}] \).
  This equation is provable in Bounded RI, but requires an additional
  lemma (a generalization of the equation):
  \( \recup\ \prefix{*}\ i\ (x - \symb{1})\ x \approx \uprec\ i\ x\ [x \geq 1] \).
  Fortunately, this lemma can be found in a systematic way, using the
  technique from \cite{fuh:kop:nis:17}.
\item Right side: here, we obtain the proof goal
  \( \downloop\ (x - \symb{1})\ x \approx \taildown\ \Qsymb\ \symb{1}\ 
  (x - \symb{1})\ x\ [x \geq \symb{1}] \).
  Again, we \emph{can} prove this, but it does require the generalization
  of the proof goal to
  \( \downloop\ i\ x \approx \taildown\ \Qsymb\ \symb{1}\ i\ x \).
  This generalization can be found using the systematic lemma generation
  technique from \cite{hag:kop:23}.
  (Note that we could also succeed with a lemma
  \( \downloop\ i\ x \approx \taildown\ \Qsymb\ \symb{1}\ i\ x\ [x \geq
  \symb{1} \wedge i \geq \symb{0}] \), obtained from the proof goal simply
  by dropping the requirement that \( i = x - 1 \).)
\end{itemize}
 
Thus, although we are not immediately done by using recursor equivalences,
combining them with existing lemma generation approaches allows us to
prove equivalences that were out of reach before.
\end{example}

\subsection{Two-sided matching}
\label{sec:twosided-matching}

In two-sided matching, we simultaneously apply template-recursor
equivalences on \emph{both} sides of an equation, and can therefore do
the subsequent RI reasoning entirely in the setting of the recursors.
Doing this, we often end up with a proof goal that is an \emph{instance} of
a recursor equivalence, and can therefore be concluded in a single step
(essentially: applying the recursor equivalence yields a remaining proof
goal which is identical terms on both sides -- meaning that there is no
remaining proof obligation).

\begin{example}
\label{ex:twosided-matching1}
We consider again $\factTU\ x \approx \factRD\ x\ [x \geq \symb{1}]$.
As before, this equation rewrites to
$\uploop\ x\ \symb{2}\ \symb{1} \approx \factRD\ x\ [x \geq \symb{1}]$.
If we now apply template equivalences on both sides at the same time, we
obtain the proof goal
\(\tailup\ \prefix{*}\ \symb{2}\ x\ \symb{1} \approx \recdown\ \prefix{*}\ \symb{2}\ x\ \symb{1}\ [x \geq \symb{1}] \).  This is an instance of the bounded
inductive theorem
\(
\TURDlem\ 
\tailup\ \funcvar\ \lowvar\ \upvar\ \cvar
\approx 
\recdown\ \funcvar\ \lowvar\ \upvar\ \cvar
\)
from \lemref{uncond-temp-eq}, so can be immediately removed.
\end{example}

It is not obvious that either one- or two-sided matching is preferable.
However, using one over the other can lead to different lemma generation
techniques being useful.

\begin{example}\label{ex:twosided-matching2}
Consider the equation $\factTD\ x \approx \factRU\ x \ [x \ge \symb{1}]$.
Rewriting both sides as far as we can, and then applying template
equivalences on both sides at once, we obtain the proof goal
\( \taildown\ \prefix{*}\ \symb{1}\ (x-\symb{1})\ x \approx
\recup\ \Qsymb\ \symb{1}\ (x-\symb{1})\ x \), given that there is a rule
\( \Qsymb\ x\ y \to y * x \).

While this looks \emph{similar} to $\TDRUlem$, it is not an instance because the
first arguments to \( \taildown \) and \( \recup \) are not the same.
However, the shape of the equation, and the desire to see the equation as an
instance of a known inductive theorem does immediately suggest the lemmas
\( \taildown\ \prefix{*}\ i\ y\ a \approx \taildown\ \Qsymb\ i\ y\ a \) and
\( \recup\ \prefix{*}\ i\ y\ a \approx \recup\ \Qsymb\ i\ y\ a \), both of which
can be proven in a straightforward way, and which suffice to bring the
proof goal in a form that is an instance of the known equation
\(
\taildown\ \funcvar\ \lowvar\ \upvar\ \cvar
\approx
\recup\ \funcvar\ \lowvar\ \upvar\ \cvar
\).
\end{example}

\begin{remark}
Note that the reasoning in
Examples~\ref{ex:onesided-matching1}--\ref{ex:twosided-matching2}
only proves that the equations under consideration are inductive theorems.
To see that they are \emph{bounded} inductive theorems we must impose some
additional requirements on $\succ$ (e.g., that
$\tutemp[\ivar,\accvar] \succeq \tailup\ \Qsymb\ \ivar\ \upterm\ \accvar$);
and using different choices (e.g., one-sided matching versus two-sided
matching, or the side of the equation on which we apply an induction
hypothesis) yields different requirements.
Regardless of these choices, we believe that it is generally easy to
satisfy at least the ordering requirements generated by using the
template-recursor equivalences and the recursor equivalences.  This will
be discussed further in \Cref{sec:temp-RI-integration}.
\end{remark}
\myparagraph{RI integration}
The main challenge in an automatic implementation of the template method is the development of heuristics to apply one-sided matching or two-sided matching. 
As we just demonstrated, there is no unique best approach, and different strategies might lead to different outcomes. We expect this to be a matter of trial and error.

\subsection{Conditional recursor equivalence}
\label{sec:cond-temp-eq}

The remaining recursor equivalences can only be proved conditionally.  For
example, $\tailup\ \funcvar\ \lowvar\ \upvar\ \cvar \approx \taildown\ 
\funcvar\ \lowvar\ \upvar\ \cvar$ is \emph{not} an inductive theorem;
however, if $\afun :: \int \to \int \to \int$ is a commutative and
associative function symbol, then $\tailup\ \afun\ \lowvar\ \upvar\ \cvar
\approx \taildown\ \afun\ \lowvar\ \upvar\ \cvar$ is one.
Such assumptions are expressed through a set $\mathcal{A}$ of axioms:
equations that consider specific fixed terms, required to be
bounded ground convertible.

\begin{definition}[Conditional bounded inductive theorems]
\label{def:condind}
We assume given a fixed \lcstrs\ with signature $\Sig$ and rules $\rules$, and
a well-founded ordering \( \succ \) that includes \( \rw \cup \supterm \).
Fix typed constants $\square_1,\dots,\square_n$.
If $C,D,E$ are context functions in $n$ parameters, we call an expression
$C \approx D\ [E]$ a \emph{$n$-ary context equation}, and we call
$C \succ D\ [E]$ a \emph{$n$-ary context ordering requirement}.
Let $\mathcal{A}$ and $\eqs$ be sets of $n$-ary context equations and
$\mathcal{O}$ a set of $n$-ary context ordering requirements.
We say that ``$\mathcal{A} \vdashc \eqs\ \text{if}\ \mathcal{O}$'' is a
\emph{conditional bounded inductive theorem} if, for all terms $s_1,\dots,s_n$
of the appropriate types, we have:

if \( C[s_1,\dots,s_n] \approx D[s_1,\dots,s_n]\ [E[s_1,\dots,s_n]] \)
is a bounded inductive theorem
for all \( C \approx D\ [E] \in \mathcal{A} \), \\
\indent
and \( C[s_1,\dots,s_n] \succ D[s_1,\dots,s_n]\ E[s_1,\dots,s_n] \) holds for
all \( C \succ D\ [E] \in \mathcal{O} \) \\
\indent
then 
\( P[s_1,\dots,s_n] \approx Q[s_1,\dots,s_n]\ [R[s_1,\dots,s_n]] \) is a
bounded inductive theorem for all \( P \approx Q\ [R] \in \eqs \).

\noindent
(Note that ``is a bounded inductive theorem'' implicitly requires
\( C[s_1,\dots,s_n] \approx D[s_1,\dots,s_n]\ [E[s_1,\dots,s_n]] \) to be
a valid equation; e.g., if $\Box_i$ occurs inside the constraint $E$, then
$s_i$ must be a theory term.)
\end{definition}

\begin{lemma}\label{lem:condTempEq}
Let $\square_1,\square_2 :: \int \to \int \to \int$.
The following are conditional bounded inductive theorems:
\\
\(
\begin{aligned}
&\textbf{(1).}\ \  
\{
  \square_1\ x\ (\square_2\ y\ z) \approx \square_2\ (\square_1\ x\ y) \ z
    ,\ \ 
    \square_1\ x\ y \approx \square_2\ y \ x
\}
\vdash
\{ \tailup\ \square_1\ \lowvar\ \upvar\ \accvar \approx
  \taildown\ \square_2\ \lowvar\ \upvar\ \accvar
\} 
\ \ \text{if}  
\\
&
\left\{\! \!
\begin{aligned}
\tailup\ \square_1\ \lowvar\ \upvar\ \accvar & \succ \taildown\ \square_2\ \lowvar_1\ \upvar\ (\square_1\ \lowvar\ \accvar) && \! \! \! \! \!
  [\lowvar \le \upvar \wedge \lowvar_1 = \lowvar + \symb{1} \wedge \upvar_1 = \upvar - \symb{1}] \\
\taildown\ \square_2\ \lowvar_1\ \upvar\ (\square_1\ \lowvar\ \accvar) & \succ
  \taildown\ \square_2\ \lowvar_1\ \upvar_1\ (\square_1\ \lowvar\ (\square_2\ \accvar\ \upvar)) && \! \! \! \! \!
  [
    \lowvar \le \upvar 
    \wedge 
    \lowvar_1 = \lowvar + \symb{1} 
    \wedge 
    \upvar_1 = \upvar - \symb{1}
    \wedge 
    \lowvar_1 \le \upvar 
  ]
\end{aligned}
\! \right\}
\\
&\textbf{(2).}\ \  
\{
  \square_1\ x \ (\square_1\ y\ z) \approx \square_1\ y\ (\square_1\ x\ z)
  , \ \ 
  \square_1\ x\ y \approx \square_2\ y\ x \} \vdash
\{ \taildown\ \square_2\ \lowvar\ \upvar\ \accvar \approx
  \recdown\ \square_1\ \lowvar\ \upvar\ \accvar \}
\ \ \text{if} 
\\ 
&
\left\{\! \!
\begin{aligned}
\taildown\ \square_2\ \lowvar\ \upvar\ \accvar 
&\succ 
\recdown\ \square_1\ \lowvar\ \upvar_1\ (\square_2\ \accvar\ \upvar)
&& \! \!
[\lowvar \le \upvar \wedge \upvar_1=\upvar-\symb{1}]
\\
\recdown\ \square_1\ \lowvar\ \upvar_1\ (\square_2\ \accvar\ \upvar)
&\succ 
\square_1\ \upvar_1\ (\square_1\ \upvar\ (\recdown\ \square_1\ \lowvar\ \upvar_2\ \accvar))
&& \! \!
[
    \lowvar \le \upvar \wedge x \le \upvar_1 \wedge \upvar_2 = \upvar_1 - \symb{1}
]
\end{aligned}
\! \right\}
\end{aligned}
\)
\\
\(
\begin{aligned}
&\textbf{(3).}\ \
\{ \square_1 \ (\square_1\ x\ y)\ z \approx \square_1\ (\square_1\ x\ z)\ y
  , \ \ 
\square_1\ x\ y \approx \square_2\ y\ x
\} \vdash
\{ \tailup\ \square_2\ \lowvar\ \upvar\ \accvar \approx
  \recup\ \square_1\ \lowvar\ \upvar\ \accvar \}
\ \ \text{if} 
\\
&
\left\{\! \!
\begin{aligned}
\tailup\ \square_2\ \lowvar\ \upvar\ \accvar 
&\succ 
\recup\ \square_1\ \lowvar_1\ \upvar\ (\square_2\ \lowvar\ \accvar)
&& \! \!
[\lowvar \le \upvar \wedge \lowvar_1 = \lowvar+\symb{1}]
\\
\recup\ \square_1\ \lowvar_1\ \upvar\ (\square_2\ \lowvar\ \accvar)
&\succ
\square_1\ (\square_1\ (\recup\ \square_1\ \lowvar_2\ \upvar\ \accvar)\ \lowvar)\ \lowvar_1 
&& \! \!
[\lowvar \le \upvar \wedge \lowvar_1 \le \upvar \wedge \lowvar_2 = \lowvar_1 + \symb{1}] 
\end{aligned}
\! \right\}
\\
&\textbf{(4).}\ \
\{\square_1\ (\square_2\ x\ y)\ z \approx \square_2\ x\ (\square_1\ y\ z)
  , \ \ 
  \square_1\ x\ y \approx \square_2\ y\ x
\} \vdash
\{ \recup\ \square_1\ \lowvar\ \upvar\ \accvar \approx
  \recdown\ \square_2\ \lowvar\ \upvar\ \accvar \}
\ \ \text{if}
\\
&
\left\{\! \!
\begin{aligned}
\recup\ \square_1\ \lowvar\ \upvar\ \accvar 
&\succ 
\square_1\ (\recdown\ \square_2\ \lowvar_1\ \upvar\ \accvar)\ \lowvar
&& \! \!
[\lowvar \le \upvar \wedge \lowvar_1 = \lowvar + \symb{1}]
\\
\square_2\ \upvar\ (\recdown\ \square_2\ \lowvar\ \upvar_1\ \accvar)
&\succ 
\square_2\ \upvar\ (\square_1\ (\recdown\ \square_2\ \lowvar_1\ \upvar_1\ \accvar)\ \lowvar)
&& \! \!
[\lowvar \le \upvar \wedge \lowvar_1 = \lowvar + \symb{1} \wedge \upvar_1 = \upvar - \symb{1} \wedge \lowvar_1 \le \upvar]
\end{aligned}
\!
\right\}
\end{aligned}
\)
\end{lemma}

Note that if $\square_1$ and $\square_2$ in \lemref{condTempEq} are
instantiated by the same associative-commutative term, the axioms are always
satisfied, but we only require what is actually needed to prove the
equation in $\eqs$.

\medskip
With \lemref{condTempEq} we can prove the four remaining factorial equivalences.
Like in the previous examples, we either apply one-sided matching or two-sided matching.
The only difference is that we now have to establish bounded ground convertibility of the axiom
sets, which in our case involves a trivial verification as they all instantiate both $\square_1$
and $\square_2$ by $*$, or by a symbol $\Qsymb$ with $\Qsymb\ x\ y \to y * x$.
We will therefore skip the details of the proofs.

\subsection{Another example: equivalence with composite operators}
\label{sec:templateCompOp}

We have seen how the template approach works on simple, first-order examples
through the various factorial implementations.  Here, we aim to illustrate
that Lemmas~\ref{lem:uncond-temp-eq} and~\ref{lem:condTempEq} are general
enough to handle more involved examples that naturally occur in real-life
programming, including cases where the context functions are distinct higher-order constructs.
Consider below $\funfactTU$ and $\funfactTD$ which are higher-order
variants of $\factTU$ and $\factTD$, computing $(h, x) \mapsto
\prod_{i=1}^x h(i)$ for all $x \ge 1$.
\[
\begin{aligned} 
\funfactTU\ h\ x
&\to 
\uploop\ h\ x\ \symb{1}\ \symb{1}
&&
% []
&
\funfactTD\ h\ x
&\to 
\downloop\ h\ x\ \symb{1}
&&
% []
\\
\uploop\ h\ x\ i\ a 
&\to 
a 
&&
[i > x]
&
\downloop\ h\ x\ a
&\to 
a
&&
[x < \symb{1}]
\\ 
\uploop\ h\ x\ i\ a
&\to 
\uploop\ h\ x\ (i+\symb{1})\ ((h\ i)*a)
&&
[i \le x]
&
\downloop\ h\ x\ a
&\to 
\downloop\ h\ (x-\symb{1})\ (a*(h\ x))
&&
[x \ge \symb{1}]
\end{aligned}
\]

Compared to $\factTU$ and $\factTD$, the expressions in the accumulator arguments are changed: $i*a$ is replaced by $(h\ i)*a$ and $a*x$ is replaced by $a*(h\ x)$.
To match them with \Cref{tab:templates}, we choose:

\begin{minipage}[t]{0.6\textwidth}
\begin{itemize}[label = $\triangleright$]
\item Tailup: $T^+_{F,x} = \uploop\ h\ x\ \square_1\ \square_2$ with
  $F := (h\ \square_1) * \square_2$;
\item Taildown: $T^-_{G,1} = \downloop\ h\ \square_1\ \square_2$ with
  $G := \square_1 * (h\ \square_2)$.
\end{itemize}
\end{minipage}
\begin{minipage}[t]{0.3\textwidth}
\vspace{-20pt}
\[
\boxed{
\begin{aligned} 
\underline{\symb{F}}\ x_1\ x_2 
&\to
(h\ x_1)*x_2
\\
\underline{\symb{G}}\ x_1\ x_2 
&
\to x_1*(h\ x_2)
\end{aligned}}
\]
\end{minipage}

Hence, using \lemref{temp-rec-eq}, both $\funfactTU\ h\ x \approx
\tailup\ \underline{\symb{F}}\ \symb{1}\ x\ \symb{1}$ and
$\funfactTD\ h\ x \approx \taildown\ \underline{\symb{G}}\ \symb{1}\ x\ 
\symb{1}$ are bounded inductive theorems.
To prove that $\funfactTU\ h\ x \approx \funfactTD\ h\ x$ is an inductive
theorem, it therefore suffices to show that
$\tailup\ \underline{\symb{F}}\ \symb{1}\ x\ \symb{1} \approx
\taildown\ \underline{\symb{G}}\ \symb{1}\ x\ \symb{1}$ is one.
This is an instance of $\tailup\ \underline{\symb{F}}\ u\ v\ w \approx
\taildown\ \underline{\symb{G}}\ u\ v\ w$, which is a bounded inductive
theorem by \lemref{condTempEq}.$\textbf{(1)}$, provided that the
equations in the condition are, if we take $\underline{\symb{F}}$ and
$\underline{\symb{G}}$ for $\square_1$ and $\square_2$ respectively.
Since we can establish that these are indeed bounded ground convertible, we
have shown the result using two-sided matching.

\section{Integrating templates in Rewriting Induction}
\label{sec:temp-RI-integration}
In this section, we explain the technical details of how templates can be used as a lemma-generation method in Bounded RI. 
As said, formally presenting Bounded RI is beyond the scope of this paper. Therefore, rather than presenting the RI proofs, we discuss the corresponding induction proofs in a self-contained manner. 
In particular, we focus on the ordering requirements arising in the different components of template-based proofs and explain how these can be recombined in concrete instances to obtain a complete induction proof.

Nevertheless, all proofs discussed in this paper have been carried out in \cora\ and are available at~\url{https://cs.ru.nl/~cynthiakop/experiments/lsfa26/}.

\subsection{Template-recursor equivalences}\label{sec:tempreceq}
The equivalences of \lemref{temp-rec-eq} can all be established by straightforward induction proofs, each following a similar induction procedure. As a representative example, we consider the equivalence
 $\tutemp[\ivar,\accvar]
\approx \tailup\ \Qsymb\ \ivar\ \upterm\ \accvar [\strue]$:
\begin{itemize}[label=$\triangleright$]
\item we prove it for all $\ivar,\accvar$, by induction on
  $\{ \tutemp[\ivar,\accvar],\ \tailup\ \Qsymb\ \ivar\ \upterm\ \accvar \}$ using the multiset
  extension of $\succ$
\item there are two cases: $\ivar > \upterm$ and $\ivar \leq \upterm$
\item in the former case, both sides reduce to $\accvar$ and we are done
\item in the latter case, $\tutemp[\ivar,\accvar] \rw \tutemp[\ivar+\symb{1},
  \ff{\ivar}{\accvar}]$ and $\tailup\ \Qsymb\ \ivar\ \upterm\ \accvar \rw \tailup\ 
  \Qsymb\ (\ivar+\symb{1})\ \upterm\ (\ff{\ivar}{\accvar})$. This yields the multiset $\{\tutemp[\ivar+\symb{1},
  \ff{\ivar}{\accvar}], \tailup\ 
  \Qsymb\ (\ivar+\symb{1})\ \upterm\ (\ff{\ivar}{\accvar})\}$, which is an instance of the induction
  hypothesis, and since both sides have decreased in $\succ$ (since they are smaller by $\rw$,
  which is included in $\succ$), we are done
\end{itemize}

Note that we only used the decrease in $\rw$ to complete the proof, which is why no additional
ordering requirements are imposed.
The proofs of all template-recursor equivalences are similarly simple.

\subsection{Recursor equivalences}
\label{sec:RIint-receq}

The recursor equivalences of \lemref{uncond-temp-eq} are a bit more difficult than the
template-recursor equivalences because we have to use a second induction hypothesis.  For example, if it
is given that 
\( \tailup\ \funcvar\ \lowvar\ \upvar\ \cvar \succ \recdown\ \funcvar\ \lowvar'\ 
\upvar\ (\funcvar\ \lowvar\ \cvar)\ [\lowvar \leq \upvar \wedge \lowvar' = \lowvar + \symb{1}]
\), we could prove $\TURDlem\ \tailup\ \funcvar\ \lowvar\ \upvar\ \cvar
\approx 
\recdown\ \funcvar\ \lowvar\ \upvar\ \cvar$  as follows:
\begin{itemize}[label=$\triangleright$]
\item let \( \sterm = \tailup\ \funcvar\ \lowvar\ \upvar\ \cvar \) and
  \( \tterm = \recdown\ \funcvar\ \lowvar\ \upvar\ \cvar \);
  we will prove that \( \tailup\ \funcvar\ \lowvar\ \upvar\ \cvar \approx \recdown\ \funcvar\ 
  \lowvar\ \upvar\ \cvar \) for all $\funcvar,\lowvar,\upvar,\cvar$, by induction using the
  multiset extension of \( \succ \) on the multiset \( \{ \sterm,\ \tterm \} \)
\item if \( \lowvar > \upvar \) then both sides reduce to $\cvar$ and we are done, so assume
  \( \lowvar \leq \upvar \)
\item in this case, the sides reduce to
  \( \tailup\ \funcvar\ \lowvar'\ \upvar\ (\funcvar\ \lowvar\ \cvar) \) and
  \( \funcvar\ \upvar\ (\recdown\ \funcvar\ \lowvar\ \upvar'\ \cvar) \)
  respectively,
  where \( \lowvar' \) is the value corresponding to \( \lowvar+\symb{1} \) and
  \( \upvar' \) corresponds to \( \upvar-\symb{1} \)
\item by the assumptions that (a) \( \rw \) is included in \( \succ \), and (b)
  \( \sterm \succ \recdown\ \funcvar\ \lowvar'\ \upvar\ (\funcvar\ \lowvar\ \cvar)\ 
  [\lowvar \leq \upvar \wedge \lowvar' = \lowvar + \symb{1}] \),
  we can use the induction hypothesis to observe that
  \( \tailup\ \funcvar\ \lowvar'\ \upvar\ (\funcvar\ \lowvar\ \cvar) \approx
  \recdown\ \funcvar\ \lowvar'\ \upvar\ (\funcvar\ \lowvar\ \cvar)\ [\lowvar \leq \upvar
  \wedge \lowvar' = \lowvar + \symb{1}] \) is a bounded inductive theorem. Again using
  points (a) and (b), we can increase the bounds to $\sterm$, $\tterm$ and obtain \( \rconv{\tailup\ \funcvar\ \lowvar'\ \upvar\ 
  (\funcvar\ \lowvar\ \cvar)}{\recdown\ \funcvar\ \lowvar'\ \upvar\ (\funcvar\ \lowvar\ \cvar)
  }{\sterm,\tterm}\ [\lowvar \leq \upvar \wedge \lowvar' = \lowvar + \symb{1}] \)
\item by transitivity of $\rconv{}{}{\sterm,\tterm}$, it suffices to show 
  \( \rconv{\recdown\ \funcvar\ \lowvar'\ \upvar\ (\funcvar\ \lowvar\ \cvar)}{
  \funcvar\ \upvar\ (\recdown\ \funcvar\ \lowvar\ \upvar'\ \cvar)}{\sterm,\tterm}\ [
  \lowvar \leq \upvar \wedge \lowvar' = \lowvar + \symb{1} \wedge \upvar' = \upvar - \symb{1}] \), using a second induction, on
  \( \{ \recdown\ \funcvar\ \lowvar'\ \upvar\ (\funcvar\ \lowvar\ \cvar),\ 
  \funcvar\ \upvar\ (\recdown\ \funcvar\ \lowvar\ \upvar'\ \cvar) \} \)
\item if \( \lowvar' > \upvar \), which is equivalent to saying \( \lowvar > \upvar' \), then both
  sides reduce to \( \funcvar\ \upvar\ \cvar \); so assume \( \lowvar' \leq \upvar \)
\item then the sides reduce to \( \funcvar\ \upvar\ (\underline{\recdown\ \funcvar\ \lowvar'\ 
  \upvar'\ (\funcvar\ \lowvar\ \cvar)}) \) and
  \( \funcvar\ \upvar\ (\underline{\funcvar\ \upvar'\ (\recdown\ \funcvar\ \lowvar\ \upvar''\ 
  \cvar)}) \) respectively, where \( \upvar'' \) is the value corresponding to \( \upvar' -
  \symb{1} \); applying the induction hypothesis to the underlined part, we are done!
\end{itemize}
The proof of \( \TDRUlem\ \taildown\ \lowvar\ \upvar\ \cvar \approx \recup\ \lowvar\ \upvar\ \cvar \) is
structured in the same way.

\subsection{Conditional recursor equivalences}
In the proof of 
\Cref{sec:RIint-receq},
the induction hypothesis 
\( \tailup\ \funcvar\ \lowvar'\ \upvar\ 
(\funcvar\ \lowvar\ \cvar)\approx \recdown\ \funcvar\ \lowvar'\ \upvar\ (\funcvar\ \lowvar\ \cvar)\ \break [\lowvar \leq \upvar \wedge \lowvar' = \lowvar + \symb{1}] \) was applied on the proof goal \( \tailup\ \funcvar\ \lowvar'\ \upvar\ (\funcvar\ \lowvar\ \cvar) \approx \funcvar\ \upvar\ (\recdown\ \funcvar\ \lowvar\ \upvar'\ \cvar) \) to eliminate the $\tailup$-dependency,
and continue with a proof goal where both sides depended on the same
recursor \( \recdown \). This let us start a second induction on
$\textbf{(H)}$ 
\( \recdown\ \funcvar\ \lowvar'\ \upvar\ (\funcvar\ \lowvar\ \cvar)
  \approx
  \funcvar\ \upvar\ (\recdown\ \funcvar\ \lowvar\ \upvar'\ \cvar)
  \ 
  [\lowvar \leq \upvar \wedge \lowvar' = \lowvar + \symb{1} \wedge \upvar' = \upvar - \symb{1}]
\), resulting in an equation $\textbf{(E)}$ which is a literal instance of
$\textbf{(H)}$, thus finishing the proof.
  
The proofs of the conditional recursor equivalences start similarly: the first induction hypothesis is used to obtain a resulting proof goal that only involves one recursor. 
However, the moment we want to apply the second induction hypothesis the proof procedure starts to deviate: $\textbf{(E)}$ is not a literal instance of $\textbf{(H)}$ and we first have to apply an axiom to make sure that their term shapes coincide.

Let us consider 
\(
\{
  \square_1\ x\ (\square_2\ y\ z) \approx \square_2\ (\square_1\ x\ y) \ z
    ,\ \ 
    \square_1\ x\ y \approx \square_2\ y \ x
\}
\vdash
\{ \tailup\ \square_1\ \lowvar\ \upvar\ \accvar \approx
  \taildown\ \square_2\ \lowvar\ \upvar\ \accvar
\}
\): 
\begin{itemize}[label = $\triangleright$]
\item Let $\sterm = \tailup\ \square_1\ \lowvar\ \upvar\ \accvar$ and $\tterm = \taildown\ \square_2\ \lowvar\ \upvar\ \accvar$
\item The proof starts similar as in \Cref{sec:RIint-receq}, using the
  IH to remove the $\tailup$-dependency, which requires the assumption
  $\tailup\ \square_1\ \lowvar\ \upvar\ \accvar \succ \taildown\ \square_2\ \lowvar_1\ \upvar\ (\square_1\ \lowvar\ \accvar)\ 
  [\lowvar \le \upvar \wedge \lowvar_1 = \lowvar + \symb{1} \wedge \upvar_1 = \upvar - \symb{1}]$. 
  \item We conclude that we should prove $\rconv{\taildown\ \square_2\ \lowvar_1\ \upvar\ (\square_1\ \lowvar\ \accvar)}{\taildown\ \square_2\ \lowvar\ \upvar_1\ (\square_2\ \accvar\ \upvar)}{\sterm, \tterm}$ $[\lowvar \le \upvar \wedge \lowvar_1 = \lowvar + \symb{1}$$ \wedge \upvar_1 = \upvar - \symb{1}]$. We do so by 
  induction on the multiset
  corresponding to the equation
  \[
  \textbf{(H).} 
  \quad 
  \taildown\ \square_2\ \lowvar_1\ \upvar\ (\square_1\ \lowvar\ \accvar) \approx \taildown\ \square_2\ \lowvar\ \upvar_1\ (\square_2\ \accvar\ \upvar)
  \]
  \begin{enumerate}[label = (\arabic*). ]
    \item The case $\lowvar_1>\upvar$ leads to the proof goal $\square_1\ \lowvar\ \accvar \approx \square_2\ \accvar\ \upvar\ [\lowvar \le \upvar \wedge \lowvar_1 = \lowvar + \symb{1} \wedge \upvar_1 = \upvar - \symb{1} \wedge \lowvar_1$$> \upvar]$, which we remove by our axiom $\square_1\ x\ y \approx \square_2\ y \ x$ (note that the constraint implies $x=y$). 
    \item The case $\lowvar_1 \le \upvar$ leads to the proof goal 
    \[
    (\textbf{E}). \quad 
    \begin{aligned}
    &
    \taildown\ \square_2\  \lowvar_1\ \upvar_1\ (\square_2\ (\square_1\ \lowvar\ \accvar)\ \upvar) 
    \approx 
    \taildown\ \square_2\ \lowvar\ \upvar_2\ (\square_2\ (\square_2\ a\ \upvar)\ \upvar_1)\\
    &
    [\lowvar \le \upvar \wedge \lowvar_1 = \lowvar + \symb{1} \wedge \upvar_1 = \upvar - \symb{1} \wedge \lowvar_1 \le \upvar \wedge \upvar_2 = \upvar_1 - \symb{1}]
    \end{aligned}
    \]
  \end{enumerate}
\item We cannot immediately apply the induction hypothesis $\textbf{(H)}$
  to eliminate $\textbf{(E)}$: matching the left-hand side fails due to
  the accumulator-argument (in $\textbf{(E)}$ this accumulator is headed
  by $\square_2$, while in $\textbf{(H)}$ it is headed by $\square_1$).
  Instead, we apply the axiom $\square_1\ x\ (\square_2\ y\ z) \approx \square_2\ (\square_1\ x\ y) \ z$ to obtain 
\[
    (\textbf{E}'). \quad 
    \begin{aligned}
    &
    \taildown\ \square_2\  \lowvar_1\ \upvar_1\ (\square_1\ \lowvar\ (\square_2\ \accvar\ \upvar)) 
    \approx 
    \taildown\ \square_2\ \lowvar\ \upvar_2\ (\square_2\ (\square_2\ a\ \upvar)\ \upvar_1)\\
    &
    [\lowvar \le \upvar \wedge \lowvar_1 = \lowvar + \symb{1} \wedge \upvar_1 = \upvar - \symb{1} \wedge \lowvar_1 \le \upvar \wedge \upvar_2 = \upvar_1 - \symb{1}]
    \end{aligned}
    \]
We remark that this step is only allowed if $\taildown\ \square_2\ \lowvar_1\ \upvar\ (\square_1\ \lowvar\ \accvar) \succ
  \taildown\ \square_2\ \lowvar_1\ \upvar_1\ (\square_1\ \lowvar\ (\square_2\ \accvar\ \upvar)) \break 
  [
    \lowvar \le \upvar 
    \wedge 
    \lowvar_1 = \lowvar + \symb{1} 
    \wedge 
    \upvar_1 = \upvar - \symb{1}
    \wedge 
    \lowvar_1\le \upvar 
  ]$ is satisfied. 
\item The proof is finished by applying $\textbf{(H)}$ to $\textbf{(E')}$
\end{itemize}
All the proofs for the conditional bounded inductive theorems use the same
structure.
  
\subsection{Templates as lemma generation method}
\label{sec:template-lemma-gen}
Bringing all this together, let us discuss two technically relevant aspects that need to be addressed in order to use templates as a lemma-generation method within Bounded RI.

\myparagraph{Bounded ground convertibility in $\rules$}
A key issue is that Bounded RI establishes bounded ground convertibility with respect to a given term rewriting system $\rules$, which typically does not contain the recursors from~\Cref{sec:templates}. 
However, these recursors play a crucial role in the inductive reasoning underlying the template method.
For example, trying to prove an equation like \( \factTU\ x \approx
\factRD\ x\ [x \ge \symb{1}] \), we first have to indentify one or both programs as recursor instances: \( \factRD\ x \approx \recdown\ [*]\ \symb{2}\ x\ \symb{1} \)
and
\( \tailup\ \funcvar\ \avar\ \bvar\ \cvar \approx \recdown\ \funcvar\ 
\avar\ \bvar\ \cvar \).  
To facilitate this reasoning within Bounded RI, we have to extend the rewrite system to $\rules \cup \{\tur, \tub, \rur, \rub\}$. 

Note that Lemmas~\ref{lem:temp-rec-eq}, \ref{lem:uncond-temp-eq} and
\ref{lem:condTempEq} assume that the recursor rules are included in
\( \rules \).  If this is not initially the case, we must import them and consequently establish convertibility in \( \rules \cup \rulesrec \). Thus, we obtain \( \openconv{s\gamma}{t\gamma}{s\gamma,t\gamma}{\rules \cup
\rulesrec} \) rather than \( \rconv{s\gamma}{t\gamma}{s\gamma,t\gamma} \).
Fortunately, in practice Bounded RI is typically applied in settings where ground confluence is either a given property, or can be established (in~\cite{hag:kop:26} we showed how to combine critical pairs with Bounded RI to prove ground confluence).  
With this assumption 
\( \openconv{s\gamma}{t\gamma}{s\gamma,t\gamma}{\rules \cup \rulesrec} \)
implies that there is some \( u \) such that both
\( s\gamma \to_{\rules \cup \rulesrec} u \) and
\( t\gamma \to_{\rules \cup \rulesrec} u \).  If the symbols
\( \tailup,\taildown,\recup,\recdown \) do not occur in \( s\gamma \) or
\( t\gamma \), then only rules in \( \rules \) can be applied in these
derivations, so we really have
\( s\gamma \rightarrow_\rules u \leftarrow_\rules t\gamma \): the
equation is also a bounded inductive theorem in \( \rules \).

\myparagraph{Ordering requirements}
Regarding the ordering $\succ$, we claim that the ordering requirements
imposed by $\rulesrec$ and the lemmas in this paper are pretty light: the
rules of $\rulesrec$ do not depend on those in $\rules$, and in the lemmas
the symbols in $\rules$ dominate those in $\rulesrec$.  For example, all
explicit ordering requirements, and the requirement that
\( \to_{\rulesrec}\;\subseteq\;\succ \),
are captured by a simple higher-order recursive path
ordering~\cite{guo:kop:24} with $\afun > \tailup > \taildown >
\recup > \recdown$, where $\afun$ is any symbol defining the $T$ and $R$
rules. (There is one disclaimer: for the ordering requirements in
\lemref{condTempEq} this also depends on the choice for $\square_1$
and $\square_2$; RPO-orientability certainly holds if $\square_1 =
\square_2$, or if these are non-recursive symbols like $\Qsymb$.)
Or if we use static dependency pairs~\cite{guo:hag:kop:val:24} to prove
termination of a corresponding rewrite relation as done
in~\cite{hag:kop:26}, the dependency pairs for the lemmas and the rules
in \( \rulesrec \) are easily eliminated by the dependency graph and the
value criterion. (Again, this does depend on what is used for
$\square_1$ and $\square_2$.)

If the ordering requirements do cause difficulties, it is worth noting that
the method does admit for some flexibility; for example, by applying an
axiom on the other side of the equation, different ordering requirements
arise, and also in \lemref{condTempEq} we could obtain different
requirements by applying the lemmas slightly differently, if this is
helpful for the given choice of $\square_1$ and $\square_2$.

\section{Conclusion}

In this paper, we have introduced several useful templates and showed how they can be incorporated into Bounded RI as a lemma generation method to prove program equivalence in a manual or automatic process.
In particular, templates can be seen as a generalization method operating at the level of programing constructs. 
The advantage of this approach becomes apparent in settings where existing lemma generation methods from constrained rewriting fail due to the need for non-polynomial invariants. 

We believe that the template approach lends itself well towards automation in \cora,
although we have not done so yet since our implementation of Bounded RI is
also not yet fully automatic even without adding postulates.  Once the
technical prerequisites are in place, it should not be hard to recognise
instances of the templates (as this is mostly syntactic), but the primary
challenge will be to decide how to apply the lemmas: as we have seen in
Sections~\ref{sec:onesided-matching} and \ref{sec:twosided-matching}, there
is no unique best approach, and different strategies might lead to
different outcomes.  We expect this to be a matter of trial and error.

\myparagraph{Related work}
In \Cref{sec:motivation}, we discussed the related work in~\cite{chi:aot:toy:10}, as a motivation for our RI-based template method in \lcstrss. 
We will now explain that, although their method uses a different approach to establish equivalence, the underlying technical prerequisites are similar to our approach.

The approach in~\cite{chi:aot:toy:10} is not based on well-founded induction. 
Instead, equivalence is proven via 
\emph{equivalent term rewriting systems}, using specifically designed transformation rules.  
Correctness of these transformations relies on \emph{confluence}. 
Additionally, the approach assumes \emph{sufficient completeness}, a property closely related to quasi-reductivity. 
However, the authors note that verifying sufficient completeness in practice requires termination.  
Thus, the prerequisites are similar to our approach: RI also requires
quasi-reductivity and termination. Moreover, as observed in
\Cref{sec:template-lemma-gen}, ground confluence is needed to obtain an
inductive theorem over $\rw$ rather than just
$\to_{\rules \cup \rulesrec}$.

\subsection{Future work}
The templates in~\cref{tab:templates} only capture a limited class of programs.
We aim to extend our collection of templates and corresponding equivalences.
We consider some natural candidates. 

\myparagraph{Foldl and foldr}
As explained in \cref{sec:motivation}, $\symb{foldl}$ and $\mathsf{foldr}$ may serve as recursors to represent tail recursion and non-tail recursion on foldable data structures. 
Consider another list-summation program:
\[
\symb{sumRec}\ [] \to \symb{0} \qquad \qquad \qquad 
\symb{sumRec}\ (x : xs) \to x + (\symb{sumRec}\ xs)
\]
We may use $\symb{foldl}$ and $\mathsf{foldr}$ to prove $\rconv{\symb{sum}\ l}{\symb{sumRec}\ l}{}$. 
In \cref{sec:motivation} we already asserted $\rconv{\symb{sum}\ l}{\symb{foldl}\ \prefix{+}\ \symb{0}\ l}{}$. 
Similarly, we can establish $\rconv{\symb{sumRec}\ l}{\symb{foldr}\ \prefix{+}\ \symb{0}\ l}{}$. 
Then we can complete the proof by applying the following bounded conditional inductive theorem with two-sided matching:\\
\[
\begin{aligned}
&
\{\square_1\ x\ y \approx \symb\square_2\ y\ x, \ \ 
\square_1\ (\square_1\ x\ y)\ z \approx \square_1\ (\square_1\ x\ z)\ y\} 
\vdash 
\{\symb{foldl}\ \square_1\ a\ l \approx \symb{foldr}\ \square_2\ a\ l\}\ \ \text{if}\\
&
\left\{
\begin{aligned}
\symb{foldr}\ \square_2\ a\ (x_1: l_1) \succ \square_2\ x_1\ (\symb{foldl}\ \square_1\ a\ l_1),  \ 
\symb{foldl}\ \square_1\ (\square_1\ a\ x_1)\ (x_2 : l_2) \succ \symb{foldl}\ \square_1\ (\square_1\ (\square_1\ a\ x_2)\ x_1)\ l_2
\end{aligned}
\right\}
\end{aligned}
\]

\myparagraph{Extensions and variations on $\rulesrec$}
We may define extensions and variations of $\rulesrec$. A useful extension is to generalize the step-size to arbitrary $k>0$. 
For example, we may consider 
\[
\begin{aligned}
\tailup\ \funcvar\ k\ \ivar\ \upvar\ \accvar & \to \accvar
 && [\ivar > \upvar] 
 &\qquad 
\tailup\ \funcvar\ k\ \ivar\ \upvar\ \accvar & \to
  \tailup\ \funcvar\ k\ (\ivar+k)\ \upvar\ (\funcvar\ \ivar\ \accvar)
  && [k>\symb{0} \wedge \ivar \leq \upvar] \\
\recdown\ \funcvar\ k\ \lowvar\ \ivar\ \basevar\ & \to \basevar
 && [\ivar < \lowvar] &
\recdown\ \funcvar\ k\ \lowvar\ \ivar\ \basevar & \to
  \funcvar\ \ivar\ (\recdown\ \funcvar\ k\ \lowvar\ (\ivar -
  k)\ \basevar) && [k>\symb{0} \wedge \ivar \geq \lowvar]
\end{aligned}
\]
and establish $\tailup\ \funcvar\ k\ \lowvar\ \upvar\ \accvar \approx \recdown\ \funcvar\ k\ \lowvar\ \upvar\ \accvar\ [k>\symb{0} \wedge x \equiv y \pmod{k}]$.
This can for instance be used to prove the equivalence
$\symb{facOddTU}\ x \approx \symb{facOddRD}\ x\ [x \equiv 1 \pmod{2}]$, where  \\ 
\[
\begin{aligned}
\rules_{\symb{facOddTU}}
&= 
\{
\symb{facOddTU}\ x
\to 
\uploop\ x\ \symb{1}\ \symb{1}, \ \ \ \ \ \ 
\uploop\ x\ i\ a 
\to 
a 
\ 
[i > x],\ \ \ 
\uploop\ x\ i\ a
\to 
\uploop\ x\ (i+\symb{2})\ (i*a)
\ 
[i \le x] 
\}
\\ 
\rules_{\symb{facOddRD}}
&= 
\{
\symb{facOddRD}\ x
\to 
\symb{1}
\ 
[x \le \symb{1}], 
\ \ \
\symb{facOddRD}\ x
\to 
x*(\symb{facOddRD}\ (x-\symb{2}))
\ 
[x>\symb{1}]\}
\end{aligned}
\]
A variation could be to replace the $+k$ by $/k$. This may be useful in e.g. sorting algorithms. 

\myparagraph{Recurrence relations}
Another programming construct are recurrence relations. For example, we may define a general recursive and tail-recursive recursor to represent recurrence relations of order $2$:
\[
\begin{aligned}
&\recR\ n\ \varphi\ a_0 \ a_1 
\to 
a_0 \
[n < \symb{1}]
\qquad \qquad 
\recR\ n\ \varphi\ a_0 \ a_1
\to a_1 
\ 
[n = \symb{1}]
\\
&\recR\ n\ \varphi\ a_0 \ a_1
\to 
\varphi\ 
n\ 
(\recR\ (n-\symb{2})\ \varphi\ a_0 \ a_1)\ 
(\recR\ (n-\symb{1})\ \varphi\ a_0 \ a_1)
\  
[n > \symb{1}]
\\
&\iter\ n\ i\ \varphi\ x\ y
\to 
y\ 
[i > n]   
\qquad 
\qquad \quad \ 
\iter\ n\ i\ \varphi\ x\ y
\to 
\iter\ n\ (i+\symb{1})\ \varphi\ y\ (\varphi\ i\ x\ y)\ 
[i \le n]
\end{aligned}
\]
We may establish 
$
\recR\ n\ \varphi\ a_0\ a_1 
\approx 
\iter\ n\  \symb{2}\ \varphi\ a_0\ a_1
\ 
[n > \symb{1}]$. This could be used to prove $\symb{fibR}\ x \approx \symb{fibT}\ x$, with 
\[ 
\begin{aligned}
\rules_{\symb{fibR}}
&
=
\{\symb{fibR}\ x \to \symb{0}\ [x\le \symb{0}],\ \ 
\symb{fibR}\ x \to \symb{1}\ [x=\symb{1}],\qquad 
\symb{fibR}\ x \to (\symb{fibR}\ (x-\symb{1})) + (\symb{fibR}\ (x-\symb{2}))\ [x>\symb{1}]
\}
\\
\rules_{\symb{fibT}}
&
=
\{
\symb{fibT}\ x \to \symb{t}\ x\ \symb{1}\ \symb{0}\ \symb{1},\ \ \ 
\symb{t}\ x\ i\ a_1\ a_2 \to a_1 \ [i>x],\ \ 
\symb{t}\ x\ i\ a_1\ a_2 \to \symb{t}\ x\ (i+\symb{1})\ a_2\ (a_1+a_2) \ [i \le x]
\}
\end{aligned}
\]
\newpage 
\nocite{*}
\bibliographystyle{eptcs}
\bibliography{references}
\end{document}